\documentclass[aps,prx,twocolumn,superscriptaddress,longbibliography]{revtex4-1}

\usepackage{graphicx}
\usepackage{amssymb}
\usepackage{amsthm}
\usepackage{amsmath}
\usepackage{rotating}
\usepackage{color}
\usepackage[colorlinks,citecolor=blue,linkcolor=blue]{hyperref}
\usepackage{multirow}

\begin{document}
%==============================================================================
\title{Kinetics of Crystallization and Orientational Ordering in Dipolar Particle Systems}
\author{Xian-Qi Xu}
\affiliation{Physics Department and Key Laboratory of Polar Materials and Devices, East China Normal University, Shanghai 200241, China}
\author{Brian B. Laird}
\affiliation{Department of Chemistry, University of Kansas, Lawrence, KS 66045, USA}
\author{Jeffrey J. Hoyt}
\affiliation{Department of Materials Science and Engineering, McMaster University, Hamilton, Ontario L8S 4L7, Canada}
\author{Mark Asta}
\affiliation{Department of Materials Science and Engineering, UC Berkeley, Berkeley, CA 94720, USA}
\author{Yang Yang}
\thanks{Author to whom correspondence should be addressed: yyang@phy.ecnu.edu.cn}
\affiliation{Physics Department and Key Laboratory of Polar Materials and Devices, East China Normal University, Shanghai 200241, China}
%\date{\today}
%==============================================================
\begin{abstract}
The kinetic mechanisms underlying bottom-up assembly of colloidal particles have been widely investigated in efforts to control crystallization pathways and to direct growth into targeted superstructures for applications including photonic crystals. Current work builds on recent progress in the development of kinetic theories for crystal growth of body-centered-cubic crystals in systems with short-range inter-particle interactions, accounting for a greater diversity of crystal structures and the role of the longer-ranged interactions and orientational degrees of freedom arising in polar systems. We address the importance of orientational ordering processes in influencing crystal growth in such polar systems, thus advancing the theory beyond the treatment of the translational ordering processes considered in previous investigations.  The work employs comprehensive molecular-dynamics simulations that resolve key crystallization processes, and are used in the development of a quantitative theoretical framework based on ideas from time-dependent Ginzburg-Landau theory. The significant impact of orientational ordering on the crystallization kinetics could be potentially leveraged to achieve crystallization kinetics steering through external electric or magnetic fields. Our combined theory/simulation approach provides opportunities for future investigations of more complex crystallization kinetics.
\end{abstract}
%\keywords{Subject Areas}
\maketitle
%anisotropy and the variation in magnitudes/anisotropy of kinetic coefficients with lattice structure and inter-particle interactions (or under external fields)
%==============================================================

\noindent {\bf  INTRODUCTION}
\vspace{0.1cm}

The structural ordering kinetics in crystallization can be complicated by orientational degrees of freedom (DOF) of the building-block particles. This complexity is widely prevalent in the crystallization of colloidal particles interacting with each other via anisotropic orientational interactions, ranging from colloids with induced electric/magnetic dipole moments\cite{Sullivan06,Dillmann08,Goyal10,Crassous14}, to patchy\cite{Zhang05,Doye07,Romano11} or Janus\cite{Romano11b,Jiang14,Reinhart16} colloids. In particular, this seemingly increased complexity offers advantages over the simple atomic systems, i.e., the possibility to steer the crystallization kinetics via external (electric or magnetic) fields\cite{Ristenpart03,Sullivan06,Crassous14}. However, despite the importance of quantitatively regulating colloidal self-assembly (or crystallization) process\cite{Gasser01,Anderson02,Leunissen05,Li16} and its potential for photonic applications\cite{Baumgartl07,Hynninen07,Mao13}, currently existing theoretical framework can accurately predict crystallization kinetics involving both the translational and orientational DOF, even on a qualitative level.

%~~~~~~~~~~~~~~~
\begin{figure*} [!htb]
\includegraphics[width=6.1 in]{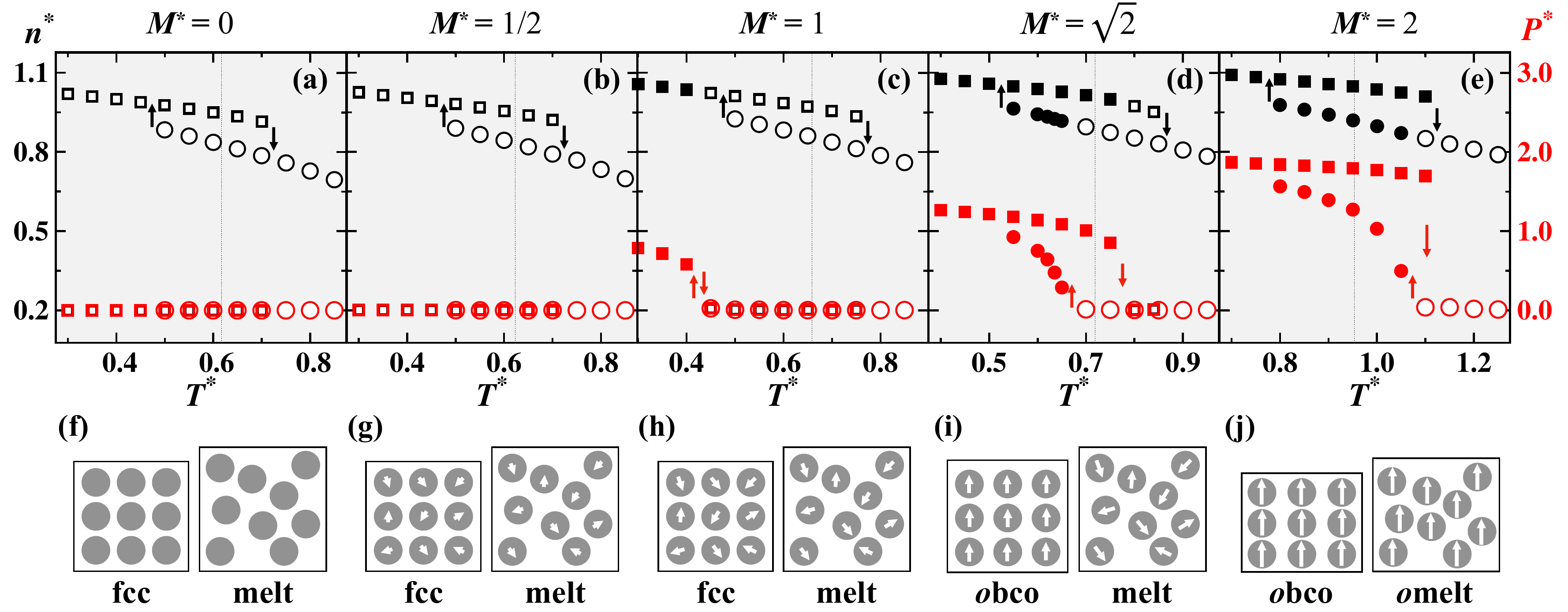}
\caption{(a-e) Heating-cooling curves in the particle number densities $n$ (black) and the polarization densities $P$ (red) corresponding to EDM particle systems with five different $M^*$. Open and filled square represent fcc and {\it o}bco structures, respectively. Open and filled circles represent orientationally disordered melt and {\it o}melt, respectively. Black and red arrows denote the crystal/melt phase transition, and Curie transition. Vertical dotted line denotes $T_\mathrm{m}$. The error bars are smaller than the size of the symbol. (f-j) Schematic diagrams of the crystal/melt phase coexistence structural type. More information about the bulk properties is summarized in the Supplementary Information.}% 
\label{fig2}
\end{figure*}
%~~~~~~~~~~~~~~~

%~~~~~~~~~~~~~~~
\begin{figure*} [!htb]
\includegraphics[width=6.2 in]{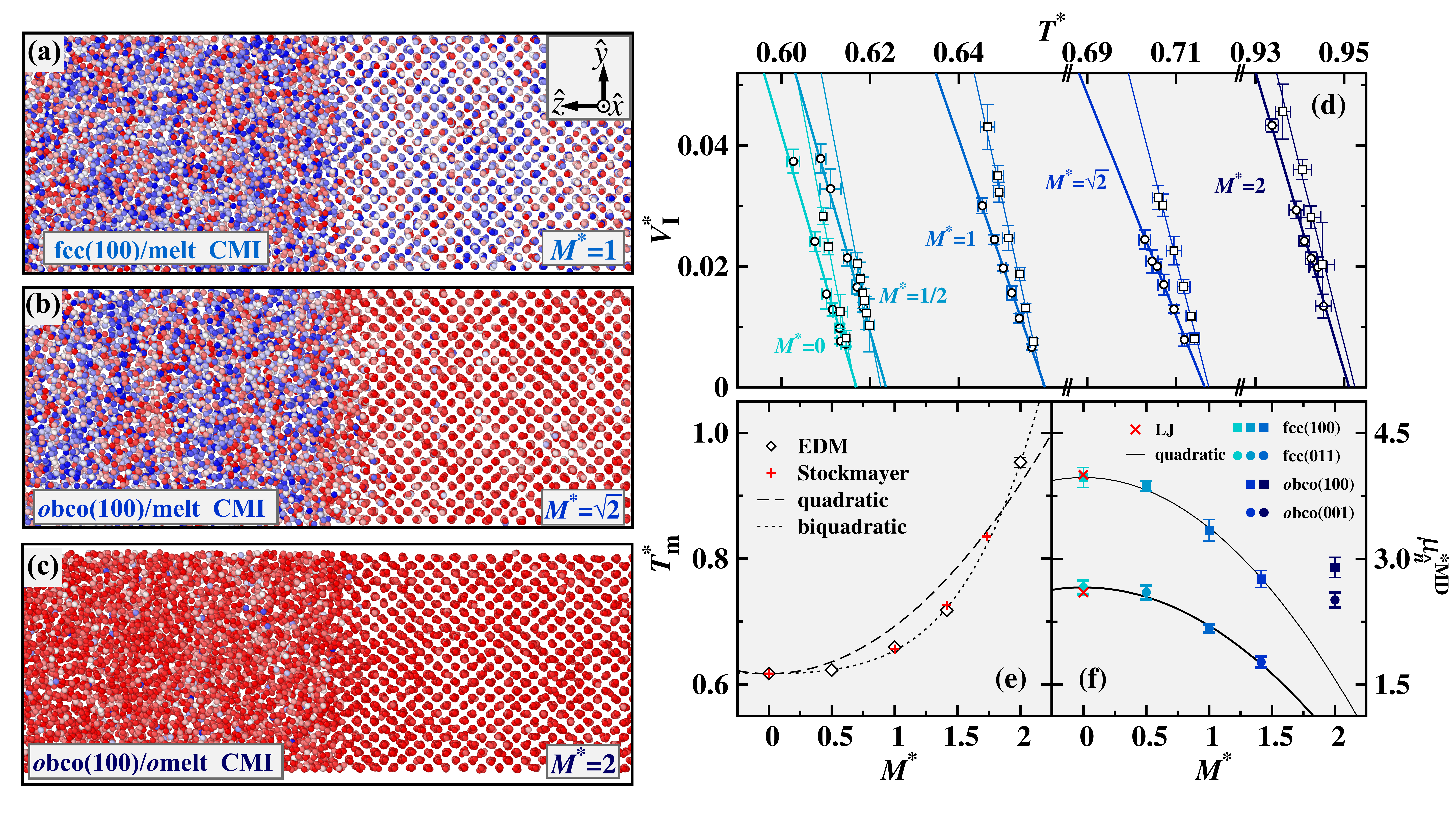}
\caption{EDM particles are color-coded as orientationally parallel (red) or antiparallel (blue) to the polarization direction, according to the degree of the particle dipolar orientation matching the mean direction of the bulk polarization. (a): the CMI is completely orientationally disordered. (b): the ferroelectric {\it o}bco crystal is grown from the orientationally disordered melt phase, in which polarization state blends from mixed color to uniform color accompanied by the increase of the degrees of crystallinity. (c): all dipolar particles are aligned in the same direction in both bulk and interface transition region. The time evolution of the NEMD trajectories is provided with the animation format in Supplementary Videos. (d): $V_\mathrm{I}$ as a function of 􏰛$T_\mathrm{I}$ derived from free-solidification simulations for the five $M^*$ systems, represented with square and circle symbols for fcc(100) (or {\it o}bco(100)) and fcc(011) (or {\it o}bco(001)). The weighted-least-square fits to the data are represented with thin and thick lines, the slopes of the fits yield the CMI kinetic coefficients $\mu$. (e): $T_\mathrm{m}$ versus $M^*$, data points (red pluses) of the Stockmayer system\cite{Wang13} coincide well with the EDM system. The biquadratic variation in $T_\mathrm{m}(M^*)$ is likely due to the biquadratic $M^*$ dependence of the second virial coefficient\cite{Bartke08phd}. (f): CMI $\mu$ of the EDM polar particle system predicted by NEMD simulations.}% 
\label{newfig2}
\end{figure*}
%~~~~~~~~~~~~~~~ 

The classical kinetic theory of crystallization is rooted in transition-state theory, the crystal/melt interface (CMI) velocity, $V_\mathrm{I}$, is expressed as the product of a term containing the thermodynamic driving force $\Delta G(T)$ and a rate-limiting term $V_0(T)$ reflecting the kinetics of the microscopic processes of particle attachment to the growing crystal, i.e., $V_\mathrm{I}=V_0(T)\{1-\exp[\Delta G(T)/k_\mathrm{B}T]\}$. Prefactor $V_0(T)$ has been suggested to be limited by the frequency with which particles collide at crystal surfaces\cite{Coriell82} or the thermally activated diffusion in the melt phase\cite{Frenkel32}, and subject to fitting parameters. Hence, the validation of different classical kinetic theoretical models relies on the fitting performance. Besides the traditional route, the modern kinetic theory of crystallization is derived from the classical density-functional theory (DFT) of freezing and inhomogeneous liquid systems.\cite{Oxtoby92,Mikheev91} As the prominent representative, the time-dependent Ginzburg-Landau (TDGL) theory recently developed by Wu et al.\cite{Wu15} suggests that the crystalline orientation-dependent density waves govern $ V_\mathrm{I}$ near melting point ($T_\mathrm{m}$) at CMIs as well as the relaxation time scale of density fluctuation. In contrast to the classical theory, TDGL theory are subject to zero fitting parameters, yielding a direct quantitative prediction of both the magnitude and the anisotropy of the CMI kinetic coefficient ($\mu$=$\frac{V_\mathrm{I}}{T_\mathrm{m}-T}$) - an essential determinant of the growth morphology\cite{Bragard02}. Unfortunately, the existing TDGL theory is limited to the body-centered cubic (bcc) crystal structure, treating only the translational DOF. 

The goal of this paper is to extend the framework of TDGL theory of CMI $\mu$ developed by Wu et al. to treat crystal structures beyond bcc and to take into account orientational ordering of the dipolar particles during the crystallization. The current study focuses primarily on CMI systems modeled with a simple dipolar particle model, which can mimic three different structural types of CMIs, including fcc/melt, {\it o}bco/melt and {\it o}bco/{\it o}melt (hereafter, ``{\it o}'' refers to ``orientationally ordered or ferroelectric or ferromagnetic'' and ``bco'' refers to ``body-centered-orthorhombic'', for simplicity). The formalism of TDGL theory covering the above CMIs has been carried out by deriving analytical expressions for $\mu$. We confirm the validity of the extended TDGL theory by using molecular-dynamics based ``computer experiments'' and show that our theory is sufficiently robust to quantitatively interpret the differences in the CMI structure and translational/rotational dynamics of dipolar particles into a definite difference in $\mu$. The roles of each contributing parameters in the analytical expression of $\mu$ governing the magnitude and anisotropy among different crystal faces are discussed, yielding fundamental insight into the connection between $\mu$ and inter-particle interactions, and the recognition of application potential in quantitatively tuning the crystallization kinetics. Our integration of simulation and theory could facilitate further extensions of the TDGL theory for broader categories of interface phase transition kinetics.

\vspace{0.3cm}
\noindent {\bf  RESULTS}
\vspace{0.1cm}

We focus primarily on dipolar particle systems modeled with the Extended Dipole Model\cite{Ballenegger04} (EDM). The use of EDM enables one to develop a generic understanding of the bulk and interface thermodynamics of dipolar particle systems. Nonetheless, compared with the point dipole model\cite{Bartke07,Wang13}, EDM is physically more relevant and it has been shown to accurately capture both structural and dielectric properties of real polar liquids\cite{Motevaselian18} in the coarse-grained MD simulation. In this work, we present a systematic simulation study of the dependence of CMI kinetic coefficient $\mu$ on the inter-particle interactions, i.e., $M$. Specifically, five different magnitudes of $M^*$=0, $\frac{1}{2}$, 1, $\sqrt{2}$, 2 are considered (varying $q$ while fixing the dipole elongation $d^*$=$0.15$). Hereafter, the superscript asterisk denotes dimensionless reduced units. Note that, the EMD model applies to both electric and magnetic dipoles. As shown in FIG.\ref{fig2} and in the Supplementary Information, EDM crystal and melt can coexist over five $M^*$ systems, mimicking three different structural types of CMIs (fcc/melt, {\it o}bco/melt and {\it o}bco/{\it o}melt).

FIG.\ref{newfig2}(a-c) shows representative CMI configurations during the non-equilibrium molecular dynamics (NEMD) simulation of growing fcc and {\it o}bco crystals. For each $M^*$, two CMI orientations are considered. Near $T_\mathrm{m}$, the interface velocities $V_\mathrm{I}$ vary linearly with the interface temperature $T_\mathrm{I}$ over all data-sets in FIG.\ref{newfig2}(d). The results of $\mu$ extracted through the slope of $V_\mathrm{I}$ versus $T_\mathrm{I}$ are summarized in FIG.\ref{newfig2}(f) and TABLE.\ref{tab2}. For $M^*$=0, both the magnitudes and the anisotropy are statistically identical to those calculated for the LJ system\cite{Huitema99}. In contrast to the biquadratic increase of $T_\mathrm{m}$ over all five $M^*$ systems, the quadratically decreasing trend of $\mu(M)$ is valid for the fcc/melt and {\it o}bco/melt CMIs, while the $M^*=2$ system does not follow the decreasing trend. The magnitude of $\mu$ of the (100) orientation is higher than that of fcc(011) or {\it o}bco(001). The magnitudes of the kinetic anisotropy for the three fcc/melt CMIs are unaffected by the dipole strength of the EDM particles and are identical to the value of the LJ system (1.52)\cite{Huitema99}. It is interesting to find that {\it o}bco/melt CMI system has an identical magnitude of kinetic anisotropy (1.56(8)) as fcc/melt CMI systems while the {\it o}bco/{\it o}melt CMI system shows a largely suppressed magnitude of kinetic anisotropy (1.16(6)). One would speculate that the structural differences due to particle orientational ordering cause the novel variation in $\mu$; however, to clarify such speculation, a quantitative TDGL theory is needed.

In the formalism of current TDGL theory (see Methods and Supplementary Information), we employ the second set of reciprocal lattice vectors (RLV) and multi-sets of order parameters to ensure the reasonable description of the interface free energy functionals. To incorporate the orientational DOF, we introduce the lowest order of a Landau-Ginzburg free energy for a ferroelectric system into the total free energy functional of the CMIs and a dissipative time constants (DTC) due to orientational ordering. The derived analytical expression (Eq.\ref{eq:GLmu}-\ref{eq:Asbco} in Methods) suggests that the magnitude of $\mu$ relies on $T_\mathrm{m}$ and $L$ (the latent heat of fusion per particle), it further implies that the kinetic anisotropy is governed by both the CMI orientation-dependent fluxes of GL order parameters and the corresponding DTCs. Next, $\mu$ for all CMI systems are predicted explicitly with Eq.\ref{eq:GLmu}-\ref{eq:Asbco} for using input parameters measured from equilibrium MD simulations (see Methods), and are compared to the NEMD simulation results, to address the questions as to what extent the current theory can predict $\mu(M)$ and reveal the complexity in CMI kinetics arises from the microscopic interface structure and the orientational DOF.

%~~~~~~~~~~~~~~~
\begin{figure} [!htb]
\includegraphics[width= 3.375 in]{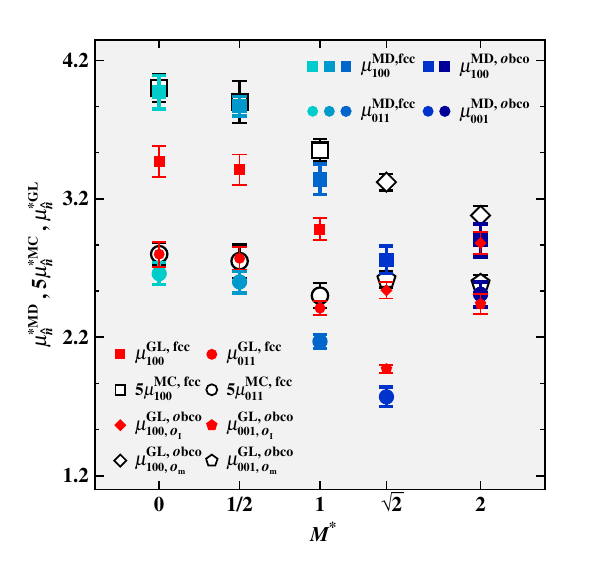}
\caption{$\mu^\mathrm{MD}$ and $\mu^\mathrm{GL}$ for the two orientations of the fcc and {\it o}bco lattice-based CMIs, as functions of $M$. Filled red squares and circles represent TDGL predictions for the fcc(100)/melt and fcc(011) /melt CMIs. Open diamonds and pentagons represent predictions for the {\it o}bco(100)/({\it o})melt and {\it o}bco(001)/({\it o})melt CMIs, determined from calculating $A_{\hat{n}}$ using integrand terms measured from the melt phase at polarization density of $o_\mathrm{m}$, while the filled red diamonds and pentagons represent predictions using integrand terms measured from the orientationally ordered melt phase at polarization density $o_\mathrm{I}$. The MC model predictions (for fcc/melt CMIs) are far from comparable with $\mu^\mathrm{MD}$, open squares and circles represent five times the data of $\mu^\mathrm{MC}$.}
\label{fig5}
\end{figure}
%~~~~~~~~~~~~~~~

%~~~~~~~~~~~~~~~
\begin{figure} [!htb]
\includegraphics[width=3.375  in]{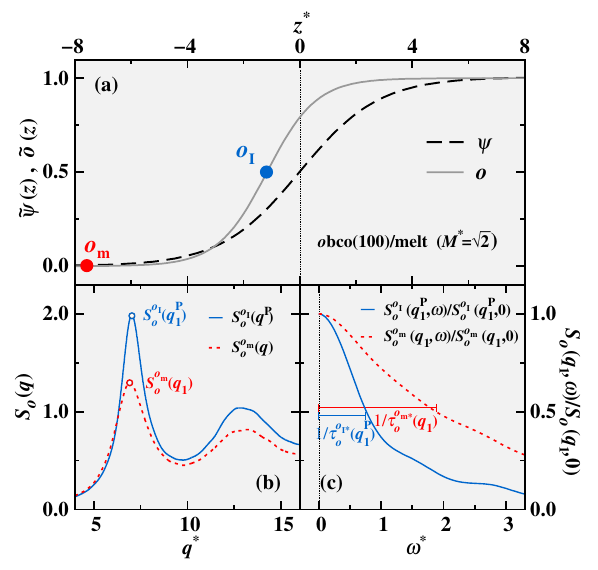}
\caption{(a) Orientational order parameter (polarization density) profile $o(z)$ defined by Eq.\ref{eq:uvopro} (solid line) and local structure order parameter profile $\psi(z)$\cite{Morris02} (dashed line) profiles of the equilibrium {\it o}bco(100)/melt CMIs. $o(z)$ and $\psi(z)$ profiles are separated by over one particle diameter, indicating a thin layer of orientationally ordered dipolar melt particles sandwiched between the isotropic bulk melt. In the vicinity of the {\it o}bco surface, frequent emergence of melt clusters in which polar particles are identified collectively aligned into the polarization orientation. The rise of these clusters does not form a complete ``wetting'' film, as seen in interfacial phase transitions\cite{Yang13}. (b) The static dipolar structure response functions for the melt phase at a polarization density of $o_\mathrm{I}$ (solid line) and the melt phase at a polarization density of $o_\mathrm{m}$ (dashed line). (c) The normalized dynamic dipolar structure response functions for the two melt phases and the corresponding orientational relaxation time scales, $\tau_{\it o}$($q_1$).}
\label{fig6}
\end{figure}
%~~~~~~~~~~~~~~~  

For the ($M^*$=0, $\frac{1}{2}$, 1) fcc/melt CMIs (top three rows of TABLE.\ref{tab2}, FIG.\ref{fig5}), TDGL theory predicts a correct anisotropy sequence $\mu^\mathrm{*GL}_{100}>\mu^\mathrm{*GL}_{011}$ for all three $M^*$. The ratio $\mu^\mathrm{*GL}_{100}/\mu^\mathrm{*GL}_{011}$ departs from $\mu^\mathrm{*MD}_{100}/\mu^\mathrm{*MD}_{011}$ by roughly 17-20\%. Despite this discrepancy, current TDGL theory significantly improves the prediction of $\mu$, comparing with its predecessor (i.e., Mikheev and Chernov (MC)\cite{Mikheev91}), which underestimates the magnitude of $\mu$ by a factor of roughly 5, see TABLE.\ref{tab2} and FIG.\ref{fig5}. We find that $1/T^2_\mathrm{m}$ is the predominant term that leads to the decreasing trends of $\mu$ in both CMI orientations with increasing $M^*$. The kinetic anisotropies $\mu^\mathrm{*GL}_{100}/\mu^\mathrm{*GL}_{011}$ are irrespective of $M^*$ within the statistical uncertainty, in agreement with NEMD predictions.

Cohen et al. show that the density wave amplitudes of the second-set RLVs decay much faster than those of the principal-set RLVs. Based on this implication, Mikheev and Chernov considered only the principal RLVs\cite{Mikheev91} in the MC model (see Supplementary Information section IV-A). Indeed, our result of density wave relaxation times is consistent with Cohen et al., e.g., $\tau^*(|\vec{K}_{\langle111\rangle}|)$ is 2.7 times greater than $\tau^*(|\vec{G}_{\langle200\rangle}|)$ for the case of $M^*=0$. (see supplementary Tab.S VI and Fig.S4). However, the two DTCs ($\varsigma^*_1$ and $\varsigma^*_2$, see Eq.\ref{eq:dtcfcc}) are found to have the identical value in TABLE.\ref{tab3}, which is a consequence that $S(|\vec{K}_{\langle111\rangle}|)$ is also around 2.7 times greater than $S(|\vec{G}_{\langle200\rangle}|)$, suggesting that the second-set RLVs should not be excluded in the TDGL theory for the non-bcc CMI systems. 

%~~~~~~~~~~~~~~~ 
\begin{table*}[hbtp]
\caption{Summary of the magnitude and anisotropy of the kinetic coefficients predicted by the NEMD simulation ($\mu_{\hat{n}}^\mathrm{MD}$), MC theory ($\mu_{\hat{n}}^\mathrm{MC}$) and the present TDGL theory ($\mu_{\hat{n}}^\mathrm{GL}$). The input parameters used in TDGL theory include melting point $T_\mathrm{m}$, latent heat $L$, and anisotropy factor $A_{\hat{n}}$. Two CMI orientations are fcc(100) and fcc(011) for $M^*$=1, $\frac{1}{2}$ and 1, and are {\it o}bco(100) and {\it o}bco(001) for $M^*=\sqrt{2}$ and 2. Subscript ``$o_\mathrm{I}$'' and ``$o_\mathrm{m}$'' denote the prediction made by employing integrand terms measured from melt phases at polarization densities of $o_\mathrm{I}$ and $o_\mathrm{m}$. Error bars represent 95\% confidence intervals on the last digit(s) shown.}
\begin{ruledtabular}
\begin{tabular}{cccccccccccc}
\multirow{2}{*}{$M^{*}$}&
\multirow{2}{*}{$\mu^\mathrm{*MD}_{100}$}&
\multirow{2}{*}{$\mu^\mathrm{*MD}_{011}$}&
\multirow{2}{*}{$\frac{\mu^\mathrm{*MD}_{100}}{\mu^\mathrm{*MD}_{011}}$}&
$\mu^\mathrm{*MC}_{100}$&
$\mu^\mathrm{*MC}_{011}$&
$\frac{\mu^\mathrm{*MC}_{100}}{\mu^\mathrm{*MC}_{011}}$&
\multicolumn{5}{c}{\;
\multirow{2}{*}{$T^{*}_\mathrm{m}$}\hspace{.35in}
\multirow{2}{*}{$L^*$}\qquad
\multirow{2}{*}{$L^*/T^{*2}_\mathrm{m}$}\quad
\multirow{2}{*}{$1/A^{*\mathrm{fcc}}_{100}$}\quad
\multirow{2}{*}{$1/A^{*\mathrm{fcc}}_{011}$}}\\
&&&&
$\mu^\mathrm{*GL}_{100}$&
$\mu^\mathrm{*GL}_{011}$&
$\frac{\mu^\mathrm{*GL}_{100}}{\mu^\mathrm{*GL}_{011}}$&
\multicolumn{5}{c}{}\\
\hline
\multirow{2}{*}{0}&
\multirow{2}{*}{3.97(12)}&
\multirow{2}{*}{2.66(8)}&
\multirow{2}{*}{1.49(6)}&
0.80(2)&0.56(1)&1.40(5)&
\multicolumn{5}{c}{
\multirow{2}{*}{0.617(1)}\quad
\multirow{2}{*}{1.017(9)}\quad
\multirow{2}{*}{2.67(3)}\quad
\multirow{2}{*}{1.30(4)}\quad
\multirow{2}{*}{1.05(3)}}\\
&&&&
3.47(11)&2.80(9)&1.24(6)&
\multicolumn{5}{c}{}\\
\multirow{2}{*}{1/2}&
\multirow{2}{*}{3.87(6)}&
\multirow{2}{*}{2.60(8)}&
\multirow{2}{*}{1.49(5)}&
0.78(3)&0.55(2)&1.41(9)&
\multicolumn{5}{c}{
\multirow{2}{*}{0.623(2)}\quad
\multirow{2}{*}{1.026(8)}\quad
\multirow{2}{*}{2.64(2)}\quad
\multirow{2}{*}{1.29(4)}\quad
\multirow{2}{*}{1.05(3)}}\\
&&&&
3.41(11)&2.77(8)&1.23(5)&
\multicolumn{5}{c}{}\\
\multirow{2}{*}{1}&
\multirow{2}{*}{3.34(13)}&
\multirow{2}{*}{2.17(5)}&
\multirow{2}{*}{1.54(7)}&
0.71(1)&0.50(1)&1.40(4)&
\multicolumn{5}{c}{
\multirow{2}{*}{0.659(1)}\quad
\multirow{2}{*}{1.070(8)}\quad
\multirow{2}{*}{2.46(2)}\quad
\multirow{2}{*}{1.21(3)}\quad
\multirow{2}{*}{0.98(2)}}\\
&&&&
2.98(8)&2.41(5)&1.24(4)&
\multicolumn{5}{c}{}\\
\hline
\multirow{2}{*}{$M^{*}$}&
\multirow{2}{*}{$\mu^\mathrm{*MD}_{100}$}&
\multirow{2}{*}{$\mu^\mathrm{*MD}_{001}$}&
\multirow{2}{*}{$\frac{\mu^\mathrm{*MD}_{100}}{\mu^\mathrm{*MD}_{001}}$}&
$\mu^\mathrm{*GL}_{100,o_\mathrm{m}}$&
$\mu^\mathrm{*GL}_{001,o_\mathrm{m}}$&
$\frac{\mu^\mathrm{*GL}_{100,o_\mathrm{m}}}{\mu^\mathrm{*GL}_{001,o_\mathrm{m}}}$&
\multicolumn{5}{c}{\qquad
\multirow{2}{*}{$T^{*}_\mathrm{m}$}\qquad\quad
\multirow{2}{*}{$L^*$}\qquad
\multirow{2}{*}{$L^*/T^{*2}_\mathrm{m}$}\quad
$1/A^{*o\mathrm{bco}}_{100,o_\mathrm{m}}$\;
$1/A^{*o\mathrm{bco}}_{001,o_\mathrm{m}}$}\\
&&&&
$\mu^\mathrm{*GL}_{100,o_\mathrm{I}}$&
$\mu^\mathrm{*GL}_{001,o_\mathrm{I}}$&
$\frac{\mu^\mathrm{*GL}_{100,o_\mathrm{I}}}{\mu^\mathrm{*GL}_{001,o_\mathrm{I}}}$&
\multicolumn{5}{c}{\qquad\qquad\qquad
\qquad\qquad\qquad\quad
$1/A^{*o\mathrm{bco}}_{100,o_\mathrm{I}}$\quad
$1/A^{*o\mathrm{bco}}_{001,o_\mathrm{I}}$}\\
\hline
\multirow{2}{*}{$\sqrt{2}$}&
\multirow{2}{*}{2.76(10)}&
\multirow{2}{*}{1.77(7)}&
\multirow{2}{*}{1.56(8)}&
3.32(6)&2.62(6)&1.27(4)&
\multicolumn{5}{c}{\multirow{2}{*}{0.718(3)}\quad\multirow{2}{*}{1.472(8)}\quad\multirow{2}{*}{2.86(2)}\quad1.16(2)\quad0.91(2)}\\
&&&&
2.54(6)&1.97(3)&1.29(4)&
\multicolumn{5}{c}{\quad\qquad\qquad
\qquad\qquad\qquad\qquad 0.89(2)\quad0.69(1)}\\
\multirow{2}{*}{2}&
\multirow{2}{*}{2.90(12)}&
\multirow{2}{*}{2.51(9)}&
\multirow{2}{*}{1.16(6)}&
3.08(7)&2.59(6)&1.19(4)&
\multicolumn{5}{c}{\multirow{2}{*}{0.953(8)}\quad\multirow{2}{*}{2.048(9)}\quad\multirow{2}{*}{2.26(2)}\quad1.36(3)\quad1.15(3)}\\
&&&&
2.88(8)&2.44(7)&1.18(5)&
\multicolumn{5}{c}{\quad\qquad\qquad
\qquad\qquad\qquad\qquad1.28(3)\quad1.08(3)}\\
\end{tabular}
\end{ruledtabular}
\label{tab2}
\end{table*}
%~~~~~~~~~~~~~~~

%~~~~~~~~~~~~~~~
\begin{table*}[hbtp]
\caption{Summary of the integrand terms in Eq.\ref{eq:Asfcc} and Eq.\ref{eq:Asbco} for the predictions of of anisotropy factor $A_{\hat{n}}$. DTC: dissipative time constant, SI of S-G: spatial integration of GL order parameter square-gradient terms. Error bars represent 95\% confidence intervals on the last digit(s) shown.}
\begin{ruledtabular}
\begin{tabular}{ccccccccccccc}
&\multicolumn{2}{c}{DTC, melt phase}&
\multicolumn{5}{c}{SI of S-G, fcc(100)/melt}&
\multicolumn{5}{c}{SI of S-G, fcc(011)/melt}\\
\hline
$M^*$&
\multicolumn{2}{c}{$\varsigma^*_1\qquad\qquad\varsigma^*_2$}&
\multicolumn{5}{c}{$8\int(\frac{\mathrm{d}u}{\mathrm{d}z})^2$\qquad$4\int(\frac{\mathrm{d}v_a}{\mathrm{d}z})^2$\ \quad$2\int(\frac{\mathrm{d}v_b}{\mathrm{d}z})^2$}&
\multicolumn{5}{c}{$4\int(\frac{\mathrm{d}u_a}{\mathrm{d}z})^2$\qquad
$4\int(\frac{\mathrm{d}u_b}{\mathrm{d}z})^2$\qquad
$4\int(\frac{\mathrm{d}v_a}{\mathrm{d}z})^2$\qquad
$2\int(\frac{\mathrm{d}v_b}{\mathrm{d}z})^2$}
\\
%\hline
0&
\multicolumn{2}{c}{0.44(1)\qquad0.44(3)}&
\multicolumn{5}{c}{1.016(8)\ \qquad0.536(4)\qquad0.198(2)}&
\multicolumn{5}{c}{0.480(4)\ \qquad0.808(12)\qquad0.584(8)\qquad0.298(4)}\\
1/2&
\multicolumn{2}{c}{0.44(2)\qquad0.44(2)}&
\multicolumn{5}{c}{1.024(8)\ \qquad0.536(4)\qquad0.200(2)}&
\multicolumn{5}{c}{0.480(4)\ \qquad0.800(8) \qquad0.588(8)\qquad0.300(2)}\\
1&
\multicolumn{2}{c}{0.45(1)\qquad0.50(2)}&
\multicolumn{5}{c}{1.024(8)\ \qquad0.532(4)\qquad0.198(2)}&
\multicolumn{5}{c}{0.480(4)\ \qquad0.804(12)\qquad0.584(8)\qquad0.298(2)}\\
\hline
&\multicolumn{6}{c}{DTC, melt phase ($o_\mathrm{m}$)}&\multicolumn{6}{c}{DTC, melt phase ($o_\mathrm{I}$)}\\
\hline
$M^*$&
\multicolumn{6}{c}{$\varsigma_{1a}^*$\quad\qquad$\varsigma_{1b}^*$\quad\qquad$\varsigma_{2a}^*$\quad\qquad$\varsigma_{2b}^*$\quad\qquad$\varsigma_{3}^*$\quad\qquad$\Omega^*$}&
\multicolumn{6}{c}{$\varsigma_{1a}^*$\quad\qquad$\varsigma_{1b}^*$\quad\qquad$\varsigma_{2a}^*$\quad\qquad$\varsigma_{2b}^*$\quad\qquad$\varsigma_{3}^*$\quad\qquad$\Omega^*$}\\
$\sqrt{2}$&
\multicolumn{6}{c}{0.49(1)\quad0.45(2)\quad0.49(1)\quad0.49(2)\quad0.41(4)\quad0.08(1)}&
\multicolumn{6}{c}{0.48(2)\quad0.47(2)\quad0.50(1)\quad0.50(1)\quad0.65(3)\;\;2.63(14)}\\
2&
\multicolumn{6}{c}{0.42(1)\quad0.35(2)\quad0.38(1)\quad0.44(2)\quad0.17(1)\quad1.23(6)}&
\multicolumn{6}{c}{0.44(2)\quad0.35(2)\quad0.41(2)\quad0.44(1)\quad0.13(1)\quad16.7(8)}\\
\hline
&\multicolumn{6}{c}{SI of S-G, {\it o}bco(100)/({\it o})melt}
&\multicolumn{6}{c}{SI of S-G, {\it o}bco(001)/({\it o})melt}\\
\hline
$M^*$&
\multicolumn{6}{c}{
$4\int(\frac{\mathrm{d}u_a}{\mathrm{d}z})^2$\quad
$4\int(\frac{\mathrm{d}u_b}{\mathrm{d}z})^2$\quad
$4\int(\frac{\mathrm{d}v_a}{\mathrm{d}z})^2$\quad
$2\int(\frac{\mathrm{d}v_b}{\mathrm{d}z})^2$\quad
$\int(\frac{\mathrm{d}o}{\mathrm{d}z})^2$}&
\multicolumn{6}{c}{
$4\int(\frac{\mathrm{d}u_a}{\mathrm{d}z})^2$\quad
$4\int(\frac{\mathrm{d}u_b}{\mathrm{d}z})^2$\quad
$4\int(\frac{\mathrm{d}v_a}{\mathrm{d}z})^2$\quad
$2\int(\frac{\mathrm{d}v_b}{\mathrm{d}z})^2$\quad
$\int(\frac{\mathrm{d}o}{\mathrm{d}z})^2$}\\
$\sqrt{2}$&
\multicolumn{6}{c}{0.524(4)\ \quad0.512(8)\ \quad0.556(4)\ \quad0.196(4)\ \quad0.151(2)}&
\multicolumn{6}{c}{0.496(4)\ \quad0.868(12)\ \quad0.624(8)\ \quad0.300(2)\ \quad0.201(2)}\\
2&
\multicolumn{6}{c}{0.580(8)\ \quad0.532(8)\ \quad0.564(8)\ \quad0.198(2)\ \quad0.011(1)}&
\multicolumn{6}{c}{0.540(4)\ \quad0.836(12)\ \quad0.588(4)\ \quad0.286(4)\ \quad0.013(1)}
\end{tabular}
\end{ruledtabular}
\label{tab3}
\end{table*} 
%~~~~~~~~~~~~~~~

We recognize differences exist in the 14 fluxes (S-G terms) of the order parameters between two orientations, the sizable anisotropy between (100) and (011) is because the former owns 10 ordering fluxes (out of 14), which have smaller magnitudes of the spatial integration than those owned by the latter. With increasing $M^*$, the values in all S-G terms seem to be universal for both CMI orientations, which suggests that the generic dimensionless particle packing is entropic driven. For a specific crystal-lattice based CMI type, the universal behavior of the S-G terms (of the GL order parameter profile) has been anticipated\cite{Wu15}. Nevertheless, our calculation provides the first compelling evidence. In contrast, DTCs are observed to depend on the inter-particle potential. We should note that the MC model employs a single DTC. Even though the MC model predicts $\mu^\mathrm{*MD}_{100}/\mu^\mathrm{*MD}_{011}$ better than the current TDGL theory (see TABLE.\ref{tab2}), the information of the dynamic property is left out. By contrast, the current two-mode TDGL theory captures the kinetic anisotropy with both the static structure and the dynamic properties.

Next, we compare the $\mu$ predicted by TDGL and NEMD for two ($M^*$=$\sqrt{2}$) {\it o}bco/melt CMIs. Our first attempted prediction using $A_{\hat{n}}$ calculated by the integrand terms measured from the isotropic bulk melt phase (labeled with subscript ``$o_\mathrm{m}$'', top half sub-row in the data row begins with ``$\sqrt{2}$'' in TABLE.\ref{tab2}, FIG.\ref{fig5}), ends up with the magnitudes of $\mu^{*\mathrm{GL}}_{100,{o_\mathrm{m}}}$ and $\mu^{*\mathrm{GL}}_{001,{o_\mathrm{m}}}$ overestimating NEMD data for over 20\% and 48\%, respectively.

We then examine the interfacial polarization state of this CMI. The results depicted in FIG.\ref{fig6}(a) indicates that the addition of the {\it o}bco surface layers may result in transforming the orientationally ordered interface melt into the orientationally ordered crystal, as suspected by Reinhart et al.\cite{Reinhart18} in their study of growing Janus colloidal crystal. Through inspecting MD trajectories, we identify frequent emergence of melt clusters in which EDM particles are collectively aligned into the polarization orientation in the vicinity of the {\it o}bco surface. A mean value $o_\mathrm{I}$ from the $o(z)$ profile could be extracted, as the polarization state of the interface orientationally ordered melt phase (supplementary Tab.S V). The calculated static and dynamic dipolar structure response functions (FIG.\ref{fig6}(b-c)), as well as the dielectric constant for the anisotropic melt phase at $o_\mathrm{I}$, show significant differences from the isotropic melt phase. With this knowledge, our second attempt using integrand terms measured from the orientationally ordered melt (labeled with subscript ``$o_\mathrm{I}$'', shown as the bottom half sub-row in the data row beginning with ``$\sqrt{2}$'' in TABLE.\ref{tab2}, FIG.\ref{fig5}) improves the prediction of $\mu$ significantly. $\mu^{*\mathrm{GL}}_{100,{o_\mathrm{I}}}$ underestimates $\mu^{*\mathrm{MD}}_{100}$ by 8\%, while $\mu^{*\mathrm{GL}}_{001,{o_\mathrm{I}}}$ overestimates $\mu^{*\mathrm{GL}}_{001}$ by 11\%.

To gain insights of above improvement, we compare integrand terms in TAB.\ref{tab3} for the two melt phases (at polarization densities of $o_\mathrm{m}$ and $o_\mathrm{I}$). The DTCs due to density fluctuations ($\varsigma_{1a,b}$ and $\varsigma_{2a,b}$) have nearly identical values, while $\varsigma_3$=$\tau_{\it o}(q_1)/S_{\it o}(q_1)$ due to orientational ordering shows a 60\% difference (FIG.\ref{fig6}(b-c)). It is worth noting that the coefficient $\Omega$ (see final paragraph of Methods) measured from the latter melt phase is about 33 times larger than that of the former melt phase. By employing the integrand terms of the melt phase at $o_\mathrm{I}$, the orientational ordering flux ($\int \varsigma_3\Omega(\frac{\mathrm{d}o}{\mathrm{d}z})^2\mathrm{d}z$) contributes around 23\% and 24\% to $A^{{\it o}\mathrm{bco}}_{100}$ and $A^{{\it o}\mathrm{bco}}_{001}$, respectively. The significant overestimation of the NEMD result mentioned above is led by the negligible percentage contribution of the ordering flux term measured from the isotropic melt phase at $o_\mathrm{m}$. With TDGL theory, any slight change in the kinetic anisotropy can be resolved by calculating subset anisotropies of S-G terms' spatial integrations. We observe that the orientational ordering flux with a subset anisotropy of 1.33 acts as a positive contributor to the total anisotropy (1.29).

From $M^*$=$\sqrt{2}$ {\it o}bco/melt to $M^*$=2 {\it o}bco/{\it o}melt CMI, the spatial integrations of S-G terms (TABLE.\ref{tab3}) varies in multiple ways. We find increases in the spatial integrations of S-G terms of $u_a$ profiles (both orientations) and $u_b$ profile ((100) orientation), while decreases are observed in the spatial integrations of S-G terms of $u_b$, $v_a$ and $v_b$ profiles ((001) orientation). As a result of the structural variations in both crystal (supplementary Tab.S IV) and melt, the four DTCs due to density fluctuation decrease in various degrees. One surprising finding is the extremely small magnitudes of $\int\mathrm{d}z(\frac{\mathrm{d}o}{\mathrm{d}z})^2$, for which the orientational ordering flux can only contribute $\sim$3\% to anisotropic factors. The tiny percentage contribution indicates that the orientational degree of freedom almost does not contribute to the crystallization kinetics of a CMI system with both its melt and crystal are orientationally ordered.  TDGL theory well reproduces the $\mu^\mathrm{MD}$ and the kinetic anisotropy (TABLE.\ref{tab2} and FIG.\ref{fig5}). The discrepancy between two predictions in {\it o}bco/{\it o}melt system is found much smaller than that of the fcc/melt and {\it o}bco/melt system, this discrepancy sequence differences among CMIs are likely closely related to the crystal structure. Finally, the suppression of the kinetic anisotropy for $M^*$=2 {\it o}bco/{\it o}melt system is found likely originates from the ordering flux of the order parameters $v$, which contributes weaker subset kinetic anisotropy to the total.
%==============================================================

\vspace{0.3cm}
\noindent {\bf  DISCUSSION AND CONCLUSION}
\vspace{0.1cm}

The merit of the TDGL theory lies in its ability to provide in-depth insight into free energy dissipation fluxes, in contrast to the classical kinetic theories and the NEMD simulation, which cannot. As suggested in the rewritten form of Eq.\ref{eq:GLmu}, $(k_\mathrm{B}T_\mathrm{m})V_\mathrm{I}A_{\hat{n}}=\frac{\Delta T}{T_\mathrm{m}}L$, the crystal growth velocity $V_\mathrm{I}$ (or $\mu$) is inversely proportional to the anisotropy factor $A_{\hat{n}}$ under a fixed thermodynamic driving force (free energy to be dissipated). From the current TDGL formalism, we observe that $A_{\hat{n}}$ is a key factor, which is governed cooperatively by multiple ordering fluxes. Each ordering flux is featured with a characteristic time scale (DTC) and length scale (spatial integration of the S-G terms). Any newly introduced complexity, whether due to RLVs (crystal structure) or additional degrees of freedom (such as orientation), is reflected by the addition of a new ordering flux term into $A_{\hat{n}}$. This knowledge can lead to an understanding of the origin of the kinetic anisotropy, elucidate the variation of $\mu_{\hat{n}}$ with crystal lattice structure and inter-particle interactions, and clarify the necessary ingredients for formulating a theory for perfectly predicting $\mu$.

The TDGL formalism for predicting the CMI kinetic coefficient can be readily extended to simple elemental crystal structures, such as hcp, hexagonal, etc. Benet et al. reported\cite{Benet14} that the $\mu$ for water is more than an order of magnitude lower than that of the LJ system, probably ascribed to the role of orientational degrees of freedom. However, we predict the decreases in the values of $\mu$ are less than an order of magnitude and are principally due to the changes in $T_m$ and DTCs. There appears to be no dissipative ordering flux arising from the ferroelectric alignment of the water molecules because neither liquid water nor ice shows spontaneous polarization under ambient pressure. Nevertheless, we are aware that the realistic ice/water CMIs\cite{Hayward02} exhibit a more complex fashion of orientational ordering than that of the current model dipolar system. A further quantitative elucidation for the participation of the orientational degree of freedom in the crystal growth of hexagonal ice, in the framework of TDGL theory, would be much needed.

A remaining challenge is the (around ten percent) discrepancy between the TDGL theory and the NEMD simulations, which is found previously in bcc/melt CMIs\cite{Wu15} and currently in fcc/melt CMIs as well as {\it o}bco/melt CMIs. We summarize here two clues that may shed light on the origin of the discrepancy. (1) A realistic density field of a CMI system requires a large set of RLVs, however, Wu et al.\cite{Wu16} suggested in their GL study of the equilibrium fcc/melt CMI, the higher-order RLVs barely affect the magnitude and anisotropy of the interfacial free energy. Based on the analytical expressions derived in this work, extending TDGL theory to include the third or more sets of RLVs also seems incapable of giving rise to a better prediction of $\mu$. For example, in the case of fcc/melt CMI, more dissipative flux terms into Eq.\ref{eq:Asfcc} would result in the higher value of $A_{\hat{n}}$ (lower magnitude of $\mu_{\hat{n}}$) and consequently a further widened discrepancy. (2) In the {\it o}bco/melt CMI systems, we have learned the importance of including the dynamic relaxation time scale and the dielectric permittivity of a melt phase at the polarization density, which is identical to the partially ordered interface layer adjacent to the crystal, as the rate limiting factors in the orientational ordering flux. This knowledge would naturally remind us to reassess the appropriateness of the assumption that the dynamic time scale due to density fluctuation at the CMI holds the same value of the bulk melt. Unfortunately, because capillary wave fluctuations dramatically wash out the transition of properties across a rough CMI, it is challenging to measure the density relaxation time for the interface melt phase over a length scale of a few particle diameters. Summing up, a suggested priority effort to eliminate the discrepancy between TDGL and NEMD predictions should focus on intrinsically\cite{Allen17} sampling the CMI positions and calculating the density relaxation time for the interface melt phase, without introducing higher-order RLVs.

To conclude, we have predicted the crystal/melt interface kinetic coefficient $\mu$ from NEMD simulation and TDGL theory. The study covers three structural types of interfacial systems (fcc/melt, {\it o}bco/melt and {\it o}bco/{\it o}melt) with two orientations described by the extended dipole model interactions. Through the introduction of the higher-order reciprocal lattice vectors and the ferroelectric order parameter to describe the orientational degree of freedom, we have extended the framework of TDGL theory to adapt fcc/melt and {\it o}bco/({\it o})melt interfaces. As predicted by the TDGL theory, the non-bcc CMI $\mu$ is determined by the square-gradient terms of different order parameter profiles multiply their corresponding dynamic relaxation times. The kinetic anisotropy is not simply governed by the lattice and CMI orientation static structure properties; it is also governed by the dynamic properties, differing from the prediction of the previous Mikheev-Chernov theory which considers only the contribution of the principal RLVs.

Our formalism of the crystallization kinetics theory predicts values of $\mu$ with an increasing dipole moment of the polar particles that are in good agreement with simulation results, with zero fit parameters. For the fcc/melt interface systems, the variation of the $\mu$ with respect to the particle dipole strength is found mainly to be due to the decrease of the ratio between latent heat and quadratic terms of the melting temperature. The anisotropy of the $\mu$ remains almost unaffected from changing the inter-particle dipolar interactions, because of the fixed percentage contributions in spatial integrals of the square gradient terms of Ginzburg-Landau order parameters. However, the stronger dipolar interaction significantly alters the lattice and polarization structure as well as the crystallization kinetics of the {\it o}bco structure-based crystal/melt interfaces. We have quantitatively demonstrated, for the first time, the engagement of both the translational and orientational degrees of freedom in the crystal growth kinetics, as well as the unique capability of TDGL theory in revealing percentage contributions of multiple ordering fluxes to the energy dissipation during crystallization, so that the physical origins of the kinetic anisotropy variation and the participation of the orientational degrees of freedom are well interpreted. In addition, we demonstrate the collective orientational reorganization of melt particles adjacent to the crystal. For the case of the ferroelctric-bco/melt CMIs, the TDGL theory identified the dynamic and dielectric properties of the partially ordered liquid as rate limiting factors in the crystal growth kinetics. 

The significant impact of orientational ordering on both the magnitude and anisotropy of $\mu$ as demonstrated in our work opens up new possibilities for steering crystallization kinetics through use of external fields. Different from the traditional way of imposing thermodynamic driving forces (i.e., temperature and pressure), the microscopic tuning of the dipolar orientational dissipative flux (the collective rotation dynamics and the polarization at CMIs) would usher in a new level of engineered polar particle crystal growth. Finally, the combined non-equilibrium theory/simulation approach proposed in this work could also have profound implications for the quantitative study of more complex crystallization kinetics, including i) solidification of alloys\cite{Yang11} (or under static magnetic fields\cite{Li14}), ii) crystallization of molecules with permanent dipole moments\cite{Benet14,Ectors15,Mandal16} and nano-sized crystallites with net dipole moments\cite{Tang06,Klokkenburg06}, iii) in-plane growth of two-dimensional surface (or confined) layers of crystalline silicon\cite{Buta07} or ice\cite{Du18,Murata19}, iv) crystal growth in fluid flows\cite{Peng17}, v) novel phase transitions (analogous to crystallization) of particles with multiple attributes, e.g., freezing of active polar colloids\cite{Bechinger16,Geyer19}, and deformable colloids\cite{Batista10}, or orientational ferroelastic transition of rod-like dipolar particles\cite{Sebastian20}.

\vspace{0.3cm}
\noindent {\bf  METHOD SECTION}
\vspace{0.2cm}

\noindent {\bf TDGL Theory for the Crystallization of Polar Particle.}

The formalism of TDGL theory covering the three types of the CMIs (fcc/melt, {\it o}bco/melt and {\it o}bco/{\it o}melt) in this study has been carried out by deriving analytical expressions for $\mu$ (see Supplementary Information for the complete details). The analytical expression for the $\mu$ of the given CMI orientation $\hat{n}$, derived from our formalism of TDGL theory, is given as,
\begin{equation}
\mu_{\hat{n}}=\frac{L}{k_\mathrm{B}T^2_\mathrm{m}A_{\hat{n}}},
\label{eq:GLmu}
\end{equation}
in which $A_{\hat{n}}$ is the anisotropy factor, given by
\begin{equation}
A_{\hat{n}}=\int\mathrm{d}z\Big[\varsigma_1\sum_{\vec{K}_i,u_{a,b}}(\frac{\mathrm{d}u_i}{\mathrm{d}z})^2+\varsigma_2\sum_{\vec{G}_i,v_{a,b}}(\frac{\mathrm{d}v_i}{\mathrm{d}z})^2\Big]
\label{eq:Asfcc}
\end{equation}
for the fcc/melt CMI, and the form of 
\begin{equation}
\begin{aligned}
A_{\hat{n}}=\int\mathrm{d}z\Big[&\varsigma_{1a}\sum_{\vec{K}_i,u_a}(\frac{\mathrm{d}u_i}{\mathrm{d}z})^2+\varsigma_{1b}\sum_{\vec{K}_i,u_b}(\frac{\mathrm{d}u_i}{\mathrm{d}z})^2+\\
&\varsigma_{2a}\sum_{\vec{G}_i,v_a}(\frac{\mathrm{d}v_i}{\mathrm{d}z})^2+\varsigma_{2b}\sum_{\vec{G}_i,v_b}(\frac{\mathrm{d}v_i}{\mathrm{d}z})^2+\\
&\varsigma_3\Omega(\frac{\mathrm{d}o}{\mathrm{d}z})^2\Big]
\label{eq:Asbco}
\end{aligned}
\end{equation}
for the {\it o}bco/melt and {\it o}bco/{\it o}melt CMI. Within the spatial integration, the sums run over 14 multiplication terms with different RLV subsets (see in the supplementary Tab.S I and Tab.S II) for the above three types of CMIs. One additional multiplication term is included, which arises due to the orientational DOF for the {\it o}bco/({\it o})melt CMI. In each multiplicative term, GL order parameter ($u$, $v$ or $o$) square-gradient (S-G) multiplies its relevant DTC ($\varsigma_{1a,b}$, $\varsigma_{2a,b}$ or $\varsigma_3$, defined as the ratio between the dynamic time scales of the density waves or orientational fluctuations relax at the given RLVs in the melt phases and their corresponding static structure factors, see in Eq.\ref{eq:dtcfcc}, Eq.\ref{eq:dtcbco1} or Eq.\ref{eq:dtcbco2}, respectively). Specifically, $u_i$ or $v_i$ are the GL order parameters describing density wave amplitudes correspond to the $i$th principal or secondary RLVs, $\vec{K}$ or $\vec{G}$, respectively. $o$ is the GL orientational order parameter, which has dimensions of dipole moment per unit volume. The coefficient $\Omega=\frac{1}{\varepsilon_0\chi n_0 k_\mathrm{B}|T_\mathrm{C}-T_\mathrm{m}|}$, in which $\varepsilon_0=1$ is the vacuum dielectric permittivity and the susceptibility $\chi=(\varepsilon_\mathrm{r}-1)$ is related to its relative permittivity (dielectric constant) $\varepsilon_{\textrm {r}}$, $n_0$ is the melt phase number density, and $T_\mathrm{C}$ is Curie temperature.

\vspace{0.5cm}
\noindent {\bf MD Simulation.}

MD simulations are performed using the program LAMMPS\cite{Plimpton95}. Each spherical particle carries mass ($m^*$=1) and moment of inertia of $I^*$= 0.117\cite{Ballenegger04}. The rigid bodies' equations of motion (two massless point charges and the spherical particle), including translational and rotational motions, are integrated with a time step $\Delta t^*$=0.001. The canonical (constant $NVT$) and isothermal-isobaric (constant $NpT$) MD simulations use a chain thermostatting method according to the generalized Nos\'e-Hoover, and Anderson approaches. The thermostat and barostat relaxation times are set as 1.0 and 10.0, respectively, to ensure Gaussian distributions of mean temperature and pressure with respect to the imposed values. The long-range Coulombic interactions are dealt with using the particle-particle particle-mesh solver and the tinfoil (conducting metal) boundary conditions, along with a real-space cutoff of 2.6$\sigma$ and a relative root-mean-square force error of approximately $10^{-5}$.
%\cite{Kamberaj05}  using the integrators developed by Kamberaj et al.\cite{Kamberaj05} with a time step $\Delta t^*$=0.001

We measure the temperature dependence of the particle number densities $n(T)$, polarization densities $P(T)$, per particle potential energies and enthalpies in a series of bulk $NpT$ simulations. Starting from a fcc sample of around 5000 EDM particles with $M^*$=0, $\frac{1}{2}$, 1, $\sqrt{2}$, 2, the temperature is step-by-step increased from zero to until the crystal melts (up to $T^*\sim1.2$) at $p^*$=0, and then is step-by-step decreased until the sample recrystallizes. Because the spontaneous polarization and the crystal contraction along the fcc $[01\bar{1}]$ direction is preferred at the Curie (fcc to {\it o}bco) transition\cite{Gao00} for systems with larger $M$, three axes of the simulation cell are chosen to follow fcc crystallographic orientations in the fashions of $x$=$[100]$, $y$=$[01\bar{1}]$, $z$=$[011]$ so that the cell dimension $L_y$ shrinks with the paraelectric-to-ferroelectric transition. We run a million MD steps for each temperature and use over a half-million MD steps for collecting averages (calculated using block averages) at each temperature and the statistical errors.

The NEMD simulations of the crystallization start from two-phase (crystal and melt) coexisting equilibrium states, which are prepared using the coexistence technique developed by Morris et al.\cite{Wang13}. The directions normal to the CMIs are defined as the $z$ axis, while the two orthogonal directions parallel to the CMIs are $x$ and $y$ axes. Periodic boundary conditions are employed in all three directions so that each simulation cell contains two CMIs. We consider two crystallographic orientations of the fcc lattice in $z$, with $M^*$=0, $\frac{1}{2}$, 1. For the fcc(100) CMIs, $x$, $y$ and $z$ axes follow $[0\bar{1}\bar{1}]$, $[01\bar{1}]$ and $[100]$, respectively. For the fcc(011) CMIs, $x$, $y$ and $z$ axes follow $[100]$, $[01\bar{1}]$ and $[011]$, respectively. The systems with two larger particle dipole moment ($M^*$=$\sqrt{2}$, 2), {\it o}bco are preferred over fcc near their corresponding $T_\mathrm{m}$. We also consider two crystallographic orientations of the {\it o}bco lattice in $z$, i.e., {\it o}bco(100) and {\it o}bco(001) CMIs. Note that, despite three different lattice parameters: $\mathrm{a_1}$, $\mathrm{a_2}$, and $\mathrm{a_3}$ as illustrated in the supplementary Fig.S3 in the Supplementary Information, {\it o}bco inherits some of the packing geometry of the fcc crystal. The [100], [010], and [001] crystallographic orientations of the {\it o}bco can be viewed as analogs to the $[100]$, $[01\bar{1}]$ and $[011]$(or $[0\bar{1}\bar{1}]$) fcc, respectively. The simulation cells have the three dimensions of approximately $18\sigma \times 15\sigma \times 170\sigma$, containing approximately 7,000 crystal phase and 38,000 melt phase EDM particles, and these numbers vary slightly from interface to interface.

The crystal/melt coexistence systems are equilibrated for a million MD steps, and ten different configurations (0.1 million steps apart) during such equilibration are employed as the starting configurations of the ten subsequent replica NEMD simulations of crystallization. The free solidification simulations are initiated by 􏰘instantaneously􏰀 imposing a small undercooling ($\Delta T/T_\mathrm{m}<5\%$) in an $Np_zA_{xy}T$ ensemble, at $p^*$=0. The CMI cross-section dimensions $A_{xy}$=$L_xL_y$ are adjusted according to the lattice parameters at each thermostatted undercooling temperature applied globally to the systems. The solidifying system quickly reaches steady-state growth, following a short transient period, and the simulations progressed until the distance between the two CMIs is less than around 15$\sigma$.
 
\vspace{0.5cm}
\noindent {\bf Measurement of $\mu$ from NEMD Simulations.}

The kinetic coefficients $\mu$ and the melting temperatures $T_\mathrm{m}$ for different CMIs are obtained from linear fits to the interface temperatures $T_\mathrm{I}$ dependences of interface velocities $V_\mathrm{I}$. We employ an atomic structural order parameter  method\cite{Morris02} to locate the CMI positions $\xi$ every 1000 MD steps. The interface velocities $V_\mathrm{I}$ can be extracted directly from monitoring the migration of the $\xi(t)$ during the steady-state regimes of each NEMD simulation system. It has been known that NEMD simulation of crystallization, which employs a single global thermostat, can lead to the presence of appreciable temperature gradients\cite{Yang11}. Such thermal gradients emerge as a peak in the interface temperature due to latent heat generation or absorption at the moving CMI during crystallization. The non-uniformity at the CMI region is compensated by the temperature in the bulk phases to produce an overall average temperature equal to that set by the global thermostat, implying that $\mu$ has to be extracted from the ratio between $V_\mathrm{I}$ and the interface temperature $T_\mathrm{I}$, rather than the average system temperature $T$. To extract $T_\mathrm{I}$, we compute the dynamic coarse-scaled profiles of temperature across the moving CMIs, $T(z)$, during steady-state regimes of each NEMD simulation system, and then compute the $T_\mathrm{I}$ from the $T(z)$ profiles, within a bin size corresponding to the 10-90 interface width. The statistical errors of the $T_\mathrm{I}$ and $V_\mathrm{I}$ were estimated by averaging 20 samples from 10 independent replica simulations - each containing two CMIs, and are reported with 95\% confidence levels. The reader is referred to Ref.\onlinecite{Yang11} for additional details of the determination of $T_\mathrm{I}$ and $V_\mathrm{I}$. Note that, the only measurement of $\mu$ for the growth of an orientationally ordered crystal before this work was the triblock Janus colloidal CMI system\cite{Reinhart18}.

%==============================================================
\vspace{0.5cm}
\noindent {\bf Validation of the TDGL Theory.}
%==============================================================

To validate the TDGL theory of $\mu$ for both fcc/melt and {\it o}bco/({\it o})melt CMI systems, we compare $\mu$ predicted by the analytical expressions (Eq.\ref{eq:GLmu}, Eq.\ref{eq:Asfcc} and Eq.\ref{eq:Asbco}, or Eq.S13 and Eq.S23 in the Supplementary Information) and from the NEMD measurements. We compute each parameter with statistical errors in the analytical expressions of $\mu$, and all errors are propagated to produce the errors of $\mu$. The values of the melting point $T_\mathrm{m}$ and latent heat $L$ for all five $M^*$ systems are listed in TABLE.\ref{tab2} in the main text. The anisotropy factor $A_{\hat{n}}$ is determined from the calculation of the S-G terms of the order parameter profiles and the static and dynamic response properties of the melt phases.

\vspace{0.5cm}
\noindent \textsl{Density Wave Amplitudes and Polarization Profiles}

All the order parameter ($u_i$, $v_i$, $o$) profiles as the function of $z$ are derived by aligning each of the fluctuating CMI $\xi(t)$ and computing time-spatial averages in the reference frame in the crystal/melt coexistences,
\begin{equation}
\begin{aligned}
u_i(z)&=\left|\langle\hat{u}_i[z-\xi(t),t]\rangle\right|,\\
v_i(z)&=\left|\langle\hat{v}_i[z-\xi(t),t]\rangle\right|,\\
o(z)&=\langle P[z-\xi(t),t]\rangle,
\label{eq:uvopro}
\end{aligned}
\end{equation}
the alignment process helps eliminate the artificial broadening effects due to the Brownian-like random walks of both CMI and crystal. The instantaneous density amplitudes are calculated from the Fourier transform of the instantaneous particle number density $n(\vec{r},t)$,
\begin{equation}
\begin{aligned}
\hat{u}_i(z,t)&=\frac{1}{V_z}\int_0^{L_x}\int_0^{L_y}\int_{z-\frac{\Delta z}{2}}^{{z+\frac{\Delta z}{2}}}\mathrm{d}x\mathrm{d}y\mathrm{d}z n(\vec{r},t)\exp(i\vec{K}_i\cdot\vec{r})\\
\hat{v}_i(z,t)&=\frac{1}{V_z}\int_0^{L_x}\int_0^{L_y}\int_{z-\frac{\Delta z}{2}}^{{z+\frac{\Delta z}{2}}}\mathrm{d}x\mathrm{d}y\mathrm{d}z n(\vec{r},t)\exp(i\vec{G}_i\cdot\vec{r}),
\label{eq:2uvopro}
\end{aligned}
\end{equation}
the averages run through discrete bins along $z$ using a bin size of $\Delta z^*=0.02$ and thus a volume of $V_z=L_xL_y\Delta_z$. The $z$ coordinate is measured relative to $\xi(t)$, $z^*<0$ for the melt and $z^*>0$ for the crystal. The instantaneous polarization density in Eq.\ref{eq:uvopro} is simply calculated as the total polarization in each discrete bin of $\Delta z$ divided by $V_z$,
\begin{equation}
P(z,t)=\frac{1}{V_z}\left|\sum_{i=1}^{N_{\mathrm{p}z}(t)}\vec{M}_i\right|,
\label{eq:3uvopro}
\end{equation}
where $N_{\mathrm{p}z}(t)$ is the dipolar particle number in the discrete bin at specific time $t$. 

For each CMI system with given $M^*$, order profiles are computed for different categories of order parameters, as listed in the supplementary Tab.S I and Tab.S II. The values of square-gradients and the bulk phases are then extracted from each order parameter profiles, listed in TABLE.\ref{tab3} in the main text and the supplementary Table.S VI, respectively. One hundred blocks of MD data (1000 configurations separated by ten steps in each block) are employed to determine the statistical errors. Fig.S4 in the Supplementary Information illustrate the reduced order parameter profiles (e.g., $\tilde{u}(z)=\frac{u(z)-u_\mathrm{m}}{u_\mathrm{c}-u_\mathrm{m}}$ vary across the interface from 0 to 1) of different categories for all CMI systems in this study.

\vspace{0.5cm}
\noindent \textsl{Dissipative Time Constants}

For the fcc/melt CMIs ($M^*$=0, $\frac{1}{2}$, 1), the two dissipative time constants (DTCs), 
\begin{equation}
\begin{aligned}
&\varsigma_1=\tau(|\vec{K}_{\langle111\rangle}|)/S(|\vec{K}_{\langle111\rangle}|), \\
&\varsigma_2=\tau(|\vec{G}_{\langle200\rangle}|)/S(|\vec{G}_{\langle200\rangle}|)
\end{aligned}
\label{eq:dtcfcc}
\end{equation}
are measured directly from the calculation of the static and dynamical structure factors of the melts ($T=T^*_\mathrm{m}$). $\tau$ is the density wave relaxation time measured as the inverse half-width of the dynamic structure factor at the given wavenumber values of $|\vec{K}_{\langle111\rangle}|$ and $|\vec{G}_{\langle200\rangle}|$ corresponding to the eight $[111]$ and six $[200]$ RLVs of the fcc crystals, respectively. The magnitudes of the $|\vec{K}_{\langle111\rangle}|$ and $|\vec{G}_{\langle200\rangle}|$ for each fcc systems (at the corresponding $T_\mathrm{m}$) are listed in the supplementary Tab.S IV in the Supplementary Information.

For the ($M^*$=$\sqrt{2}$, 2) {\it o}bco/({\it o})melt CMI systems, the four DTCs due to density waves ($\varsigma_{1a}$, $\varsigma_{1b}$, $\varsigma_{2a}$ and $\varsigma_{2b}$) are measured in a similar manner to the fcc/melt CMI systems from calculating the static and dynamical structure factors, at wavenumbers equal to the magnitudes of $|\vec{K}_{u_a}|$, $|\vec{K}_{u_b}|$, $|\vec{G}_{v_a}|$ and $|\vec{G}_{v_b}|$ (listed in Tab.S IV), respectively.
\begin{equation}
\begin{aligned}
\varsigma_{1a}&=\tau(|\vec{K}_{u_a}|)/S(|\vec{K}_{u_a}|), \\
\varsigma_{1b}&=\tau(|\vec{K}_{u_b}|)/S(|\vec{K}_{u_b}|), \\
\varsigma_{2a}&=\tau(|\vec{G}_{v_a}|)/S(|\vec{G}_{v_a}|), \\
\varsigma_{2b}&=\tau(|\vec{G}_{v_b}|)/S(|\vec{G}_{v_b}|) \\
\end{aligned}
\label{eq:dtcbco1}
\end{equation}

To determine the DTCs $\varsigma_3$ due to orientational ordering, we employ the generalized static dipolar structure response function and dynamic dipolar structure response function of the dipolar system (i.e., $S_{\it o}(q)$ and $S_{\it o}(q,\omega)$), akin to the static and dynamical magnetic susceptibility of a spin system. $S_{\it o}(q)$ is defined as the Fourier transform of the radial dipole-dipole correlation function,
\begin{equation}
S_{\it o}(\vec{q})=\frac{1}{N_\mathrm{p}} \Big\langle \sum_i\sum_{j}{\vec{M}}_i{\vec{M}}_j\mathrm{e}^{-i\vec{q}\cdot(\vec{r_i}-\vec{r_j})}\Big\rangle,
\label{eq:Soq}
\end{equation}
in which, $N_\mathrm{p}$ is the total number of the dipolar particles in the melts. The collective dynamics of orientational fluctuations in the dipolar particle melt is described through a generalized intermediate scattering function $F(\vec{q},t)$, defined as the spatial Fourier transform of the radial dipole-dipole time-correlation function,
\begin{equation}
\begin{aligned}
F_{\it o}(\vec{q},t)=\frac{1}{N_\mathrm{p}} \Big\langle \sum_i\sum_j{\vec{M}}_i(0){\vec{M}}_j(t)\mathrm{e}^{-i\vec{q}\cdot[\vec{r_i}(0)-\vec{r_j}(t)]}\Big\rangle.
\label{eq:Foqt}
\end{aligned}
\end{equation}
The time Fourier transform of the $F_{\it o}(\vec{q},t)$ into the frequency domain leads to the dynamic dipolar structure response function of the system,
\begin{equation}
S_{\it o}(\vec{q},\omega)=\frac{1}{2\pi}\int_{-\infty}^{+\infty}F_{\it o}(\vec{q},t)\mathrm{e}^{i\omega t}\mathrm{d}t.
\label{eq:Soqt}
\end{equation}
The DTC due to orientational ordering is measured as
\begin{equation}
\begin{aligned}
\varsigma_3&=\tau_{\it o}(q_1)/S_{\it o}(q_1),
\end{aligned}
\label{eq:dtcbco2}
\end{equation}
here $S_{\it o}(q_1)$ is the maximum of static dipolar structure response function at wavenumber of the first peak $q_1=|\vec{q_1}|$ (see in Tab.S IV), while $\tau(q_1)$ is the inverse half-width of the $S_{\it o}(q_1,\omega)$ (see in the supplementary Tab.S VI).

In the {\it o}bco/({\it o})melt CMI systems, the orientational order parameter (or polarization density) of the interface orientationally ordered melts, as indicated by the FIG.\ref{fig6}(a), are larger than those in bulk melts. A mean value of $o_\mathrm{I}=2o(0)-o_\mathrm{c}$ can be extracted from the $o(z)$ profile (see in supplementary Tab.S V). To provide quantitative evidence that the dissipation rate due to orientational ordering is limited by the dynamic properties of the interface melt phase at polarization density of $o_\mathrm{I}$, rather than those of the bulk melts. We calculate five DTCs as well as dielectric permittivity with melts at both polarization densities of $o_\mathrm{I}$ and $o_\mathrm{m}$ for the ($M^*$=$\sqrt{2}$, 2) {\it o}bco/({\it o})melt CMI systems.

The structure factors, the dipolar structure response functions, and the dielectric tensor becomes anisotropic (see in supplementary Fig.S5 and Fig.S6 in the Supplementary Information) for the ($M^*$=$\sqrt{2}$, 2) melt phases at $o_\mathrm{I}$ and $M^*$=2 bulk melt phase. Therefore we employ the components (e.g., $S(k^{\mathrm{P}})$, $S(k^{\mathrm{P}},\omega)$, $S_o(q^{\mathrm{P}})$, $S_o(q_1^{\mathrm{P}},\omega)$ and $\varepsilon^{\mathrm{P}}_\mathrm{r}$) parallel to the polarization direction in the validation of TDGL theory. Superscript ``P'' over the wavenumber $q$ denotes the component parallel to the polarization direction for the anisotropic (orientationally ordered) melt phase.

\vspace{0.5cm}
\noindent \textsl{Dielectric Permittivity and Coefficient $\Omega$}

The dielectric permittivity tensor $\varepsilon_\mathrm{r}$ is estimated with the standard dielectric fluctuation formula reported in the Ref.\onlinecite{Ballenegger04}. For the $M^*$=$\sqrt{2}$ system, the coefficient $\Omega=\frac{1}{\varepsilon_0(\varepsilon_\mathrm{r}-1) n_0 k_\mathrm{B}|T_\mathrm{C}-T_\mathrm{m}|}$, measured from the anisotropic melt phase at $o_\mathrm{I}$, is about 33 times larger than that measured from the isotropic melt. Specifically, the dielectric permittivity tensor, which lies parallel to the polarization direction, $\varepsilon^{*\mathrm{P}}_\mathrm{r}(o_\mathrm{I})$ and $\varepsilon^{*\mathrm{P}}_\mathrm{r}(o_\mathrm{m})$, has the values of 7.9(2) and 230(11), respectively. $n_0$ uses the melt phase number density under $T^*_\mathrm{m}$, $T^*_\mathrm{C}=0.780(3)$ is determined from the thermo-hysteresis MD simulations (see Fig.\ref{fig2}(d)). For the $M^*$=2 system, the coefficient $\Omega$ measured from the anisotropic melt phase at $o_\mathrm{I}$ is again over ten times larger than that measured from the bulk melt (which is also orientationally ordered or ferroelectric for $M^*=2$), $\varepsilon^{*\mathrm{P}}_\mathrm{r}(o_\mathrm{I})$ and $\varepsilon^{*\mathrm{P}}_\mathrm{r}(o_\mathrm{m})$ have the values of 1.33(1) and 5.5(1), $T^*_\mathrm{C}$ is 1.150(8).

%==============================================================
\vspace{0.3cm}
\noindent {\bf  AUTHOR INFORMATION}
\vspace{0.2cm}

\vspace{0.2cm}
\noindent {\bf Corresponding Author}

{\bf Yang Yang} - Physics Department and Key Laboratory of Polar Materials and Devices, East China Normal University, Shanghai 200241, China; Email: yyang@phy.ecnu.edu.cn

\vspace{0.2cm}
\noindent {\bf Authors}

{\bf Xian-Qi Xu} - Physics Department and Key Laboratory of Polar Materials and Devices, East China Normal University, Shanghai 200241, China

{\bf Brian B. Laird} - Department of Chemistry, University of Kansas, Lawrence, KS 66045, USA

{\bf Jeffrey J. Hoyt} - Department of Materials Science and Engineering,
McMaster University, Hamilton, Ontario L8S 4L7, Canada

{\bf Mark Asta} - Department of Materials Science and Engineering, UC Berkeley, Berkeley, CA 94720, USA

\vspace{0.5cm}
\noindent {\bf Notes}

The authors declare no competing financial interest.

\vspace{0.5cm}
\noindent {\bf ACKNOWLEDGEMENTS}

M.A. acknowledges funding from the U.S. Department of Energy, Office of Science, Office of Basic Energy Sciences, under Center for Molecularly Engineered Energy Materials, an Energy Frontier Research Center supported through [Award No. DE-SC0001342], and under the Fundamental Understanding of Transport Under Reactor Extremes (FUTURE) Energy Frontier Research Center. B.B.L. acknowledges support from the US National Science Foundation [Grant No. CHE-1465226].  Y.Y. acknowledge Fundamental Research Funds for the Central Universities, funding from the Chinese National Science Foundation [Grant No. 11874147], the Science and Technology Project of Shanghai Science and Technology Commission [Grant No. 18DZ1112700], ECNU Multifunctional Platform for Innovation(001) and a Key Project in Extreme Manufacturing from the Shanghai Municipal Government. 
\vspace{0.5cm}

%==============================================================
%\bibliographystyle{apsrev4-1}
\bibliography{ref}
%==============================================================

\newpage
%~~~~~~~~~~~~~~~
\begin{figure*} [!htb]
\includegraphics[width=6 in]{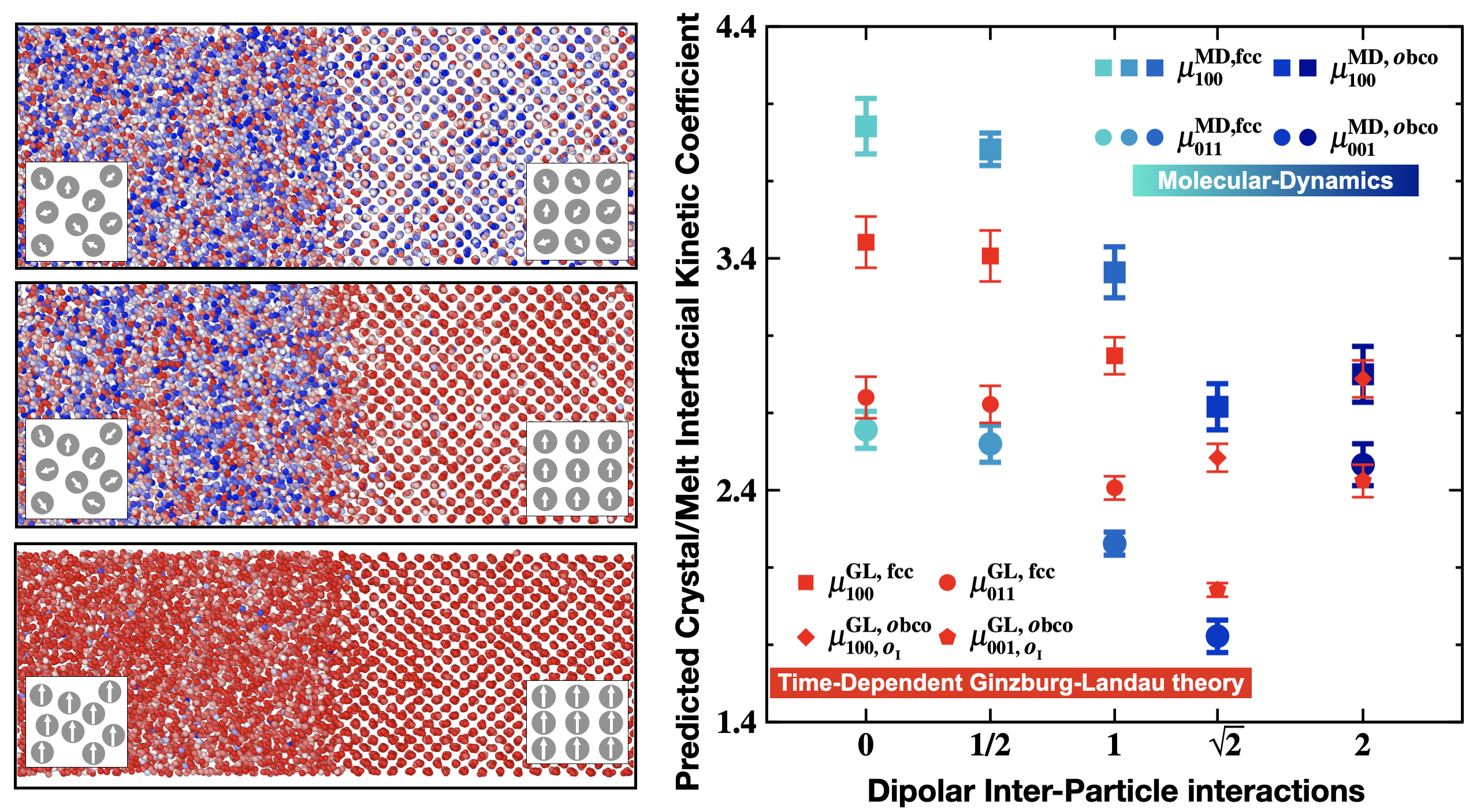}
\caption{For Table of Contents Only}
\end{figure*}
%~~~~~~~~~~~~~~~  

%==============================================================
%==============================================================

%==============================================================
\newpage

\noindent {\bf Supplementary Information: TABLE OF CONTENTS}

\vspace{0.5cm}

\noindent {\bf I. The formalism of TDGL Theory}

A. fcc/melt CMIs

B. {\it o}bco/({\it o})melt CMIs

C. Supporting Evidences

\vspace{0.4cm}
\noindent {\bf II. Supporting Information of the Extended Dipole Model}

\vspace{0.4cm}
\noindent {\bf III. Supporting Information of the MC Model Prediction of $\mu$}

\vspace{0.4cm}
\noindent {\bf IV. Supporting Data for the Validation of the TDGL Theory}

\vspace{0.4cm}
\noindent {\bf V. Supplementary Videos}

\vspace{0.4cm}
\noindent {\bf Additional References}

%==============================================================
%==============================================================
\vspace{1.5cm}
\noindent {\bf I. The formalism of TDGL Theory}
\label{sec:ap-theory}
\vspace{.3cm}
%==============================================================

In this section, we advance the TDGL theory to the case of non-bcc crystals and take into account orientational ordering of the building-block particles during crystallization, including the formalism and derivation of the analytical expressions of $\mu$ of the fcc/melt, {\it o}bco/melt and {\it o}bco/{\it o}melt CMIs.

%==============================================================
\vspace{.5cm}
\noindent A. fcc/melt CMIs
\label{ssec:ap-theory-fcc}
\vspace{.3cm}
%==============================================================

The TDGL formalism of the bcc/melt CMI by Wu et al.\cite{Wu15} is achieved through a multi-step procedure. First, the excess free energy of the CMI is represented with the free-energy functional derived initially from the DFT\cite{Haymet81,Oxtoby82}, along with the Gingzburg-Landau expansion of the free-energy density with order parameters. The GL order parameters are defined as the amplitudes of density waves corresponding to the minimal set of reciprocal lattice vectors (RLV)\cite{Shih87,Wu06}, considering the interaction between different density waves via the inclusion of higher-order terms. Secondly, one introduces the expression for the excess free energy of the non-equilibrium CMI by including the driving force of the crystallization, which is proportional to the undercooling. Thirdly, the kinetic TDGL equation is employed to describe the ordering process associated with the energy dissipation and the dynamical evolution of the order parameter field. In this picture\cite{Harrowell86,Schofield91,Oxtoby92,Shen96}, the ordering flux regarding different RLVs (in terms of the time derivative of the order parameter) is assumed to be proportional to its driving force (in terms of the local gradient of the free energy functional concerning the order parameter) via a microscopic time scale related to density fluctuations in the melt\cite{Chaikin00,Mikheev91} (hereafter referred as the dissipative time constant, DTC for simplicity). Lastly, one concludes the derivation of an analytical expression of $\mu$ by determining the solvability condition of a linear system of equations, consisting of a collection of coupled kinetic TDGL equations, corresponding to different RLV subsets.

Here, we present our detailed theoretical formulation of the TDGL theory of $\mu$ for the fcc/melt CMI, in the light of above general methodology. The whole derivation process is consists of four steps.

%\begin{equation}
%  \tag{*}
%  \int_{\partial\Omega} \omega = \int_\Omega d\omega
%  \label{eqn:Stokes}
%\end{equation}
%By \eqref{eqn:Stokes} ...

{\bf{[Step1]}}: The excess free-energy $\Delta F$ for the equilibrium fcc/melt interface relative to the free energy of the melt phase $F_\mathrm{m}$ is expressed in terms of the amplitudes of the density waves, $u_i$ and $v_i$,
\begin{equation}
 \tag{S1}
\Delta F\equiv F-F_\mathrm{m}=n_{0}k_\mathrm{B}T\int f(u_i,v_i)\mathrm{d}\vec{r}.
\label{eq:f}
\end{equation}
In the truncated expansion of the number density,
\begin{equation}
 \tag{S2}
n(\vec{r},t)=n_{0}\left[1+\sum_{\vec{K}_{i}} u_{i}(\vec{r},t) e^{i \vec{K}_{i} \cdot \vec{r}}+\sum_{\vec{G}_{i}} v_{i}(\vec{r},t) e^{i \vec{G}_{i} \cdot \vec{r}}\right],
\label{eq:denswave}
\end{equation}
$u_i$ are amplitudes of density waves (at specific time $t$) corresponding to principal RLVs $\vec{K}$ and $v_i$ are amplitudes of density waves corresponding to the second set of RLVs $\vec{G}$. The reciprocal lattice of fcc is bcc, there are eight $|{\vec{K}_{i}}|$ with identical magnitude along $\left\langle111\right\rangle$ directions and six $|{\vec{G}_{i}}|$ along $\left\langle200\right\rangle$ directions. Contributions of the third and higher RLV modes are neglected. $u_i$ and $v_i$ take the values of $u_\mathrm{c}$ and $v_\mathrm{c}$ in the fcc crystals, and the values of $u_\mathrm{m}$ and $v_\mathrm{m}$ (ideally, these two values are expected to be equal to 0) in the melt phases.

The excess free energy density $f(u_i,v_i)$ in Eq.\ref{eq:f} along the $z$ axis normal to the CMI can be expressed expressed in terms of the amplitude  $u_i$ and $v_i$ of density waves deﬁned in Eq.\ref{eq:denswave},
\begin{widetext}
\begin{equation}
 \tag{S3}
\begin{aligned} 
&2f(u_i,v_i)\approx
 a^{(2)} \sum_{i, j} c_{ij} u_{i} u_{j} \delta_{0, \vec{K}_{i}+\vec{K}_{j}}
+b^{(2)} \sum_{i, j} d_{ij} v_{i} v_{j} \delta_{0, \vec{G}_{i}+\vec{G}_{j}}
+a^{(2)}_u \sum_{i} c_{i}(\frac{\mathrm{d}u_{i}}{\mathrm{d}z})^{2}
+\sum_{i}\left[b^{(2)L}_{v}d_{i}^{L}+b^{(2)T}_{v}d_{i}^{T}\right](\frac{\mathrm{d}v_{i}}{\mathrm{d}z})^{2} \\&
-a^{(3)} \sum_{i, j, k} c_{ijk} u_{i} u_{j} v_{k} \delta_{0, \vec{K}_{i}+\vec{K}_{j}+\vec{G}_{k}}
+a^{(4)} \sum_{i, j, k, l} c_{ijkl}u_{i}u_{j}u_{k}u_{l}\delta_{0, \vec{K}_{i}+\vec{K}_{j}+\vec{K}_{k}+\vec{K}_{l}}
+b^{(4)} \sum_{i, j, k, l} d_{ijkl}u_{i}u_{j}v_{k}v_{l} \delta_{0, \vec{K}_{i}+\vec{K}_{j}+\vec{G}_{k}+\vec{G}_{l}},
\end{aligned}
\label{eq:fpower}
\end{equation}
\end{widetext}
all the numbers in parentheses in the superscripts of the multiplicative coefficients denote the order of the power series. The use of the multiplicative coefficients is convenient for normalizing the sums of the coefficients $c$'s and $d$'s to unity, e.g., $\sum_{i, j}c_{ij}\delta_{0,\vec{K}_{i}+\vec{K}_{j}}$=1, $\sum_id^{L}_i$=1, $\sum_id^{T}_i$=1 or $\sum_{i,j,k,l}d_{ijkl}\delta_{0,\vec{K}_{i}+\vec{K}_{j}+\vec{G}_{k}+\vec{G}_{l}}$=1. 

In Eq.\ref{eq:fpower}, the S-G terms arise from the spatial variation of the order parameters along $z$. Kronecker deltas ($\delta$) appear in the quadratic, cubic and quartic terms. The subscript in each $\delta$ requests closed polygons, which are made of $\vec{K}$ and $\vec{G}$ RLVs in reciprocal space. A closed polygon ensures a non-zero contribution to the functional $f$. The non-vanishing cubic term corresponds to a free-energy barrier between the crystal and melt phases and guarantees the phase transition first order\cite{Alexander78}. For the fcc/melt CMI system, the inclusion of the principal RLVs $\vec{K}$ is not enough to avoid the absence of the free-energy barrier, which is nevertheless true for the case of bcc/melt CMI system\cite{Wu15}. In addition, the closed four-sided polygons by four RLVs give rise to the two quartic terms, which ensure the stability of the crystal phase.

The multiplicative and the normalization coefficients of the quadratic terms and the S-G terms are obtained in Ref.\onlinecite{Wu16} from the comparison between the Eq.\ref{eq:fpower} and the free-energy functional expression of an inhomogeneous liquid in the formulation of the classical DFT under the assumption that the density wave decays slowly and the density wave amplitudes are constant over the scale of the atomic layer spacing\cite{Wu06}. Specifically, the coefficients for the quadratic terms are $a^{(2)}$=8/$S(|\vec{K}_{\langle111\rangle}|)$, $b^{(2)}$=6/$S(|\vec{G}_{\langle200\rangle}|)$, $c_{ij}$=1/8 and $d_{ij}$=1/6, $S(|\vec{K}|)$ and $S(|\vec{G}|)$ denote the liquid structure factors. An assumption made here relating to the RLVs in the structure factor is that the wave vectors are constants through the whole interface region, so that crystal growth velocity is not affected by the lattice spacing and density variations across the CMI\cite{Wu07}. The coefficients for the S-G terms are $a_u^{(2)}$=$-\frac{4}{3}C^{\prime\prime}(|\vec{K}_{\langle111\rangle}|)$, $c_{i}$=$\frac{3}{8}(\hat{K_{i}}\cdot\hat{n})^2$, $b_v^{(2)L}$=$-C^{\prime\prime}(|\vec{G}_{\langle200\rangle}|)$, $d_{i}^L$=$\frac{1}{2}(\hat{G_{i}}\cdot\hat{n})^2$, $b_v^{(2)T}$=$-2C^{\prime}(|\vec{G}_{\langle200\rangle}|)/|\vec{G}_{\langle200\rangle}|$ and $d_{i}^T$=$\frac{1}{4}(1-\hat{G_{i}}\cdot\hat{n})^2)$. The prime notation over the two-particle direct correlation functions $C|\vec{G}_{\langle200\rangle}|$ stands for the derivative with respect to the corresponding RLV. ``$T$" and ``$L$" in the superscriptions stand for the transverse and longitudinal component of $\hat{G}$, respectively.

Alternatively, the multiplicative and the normalization coefficients of the cubic and quartic terms are obtained\cite{Wu16} from (i) the ansatz that all the closed polygons in reciprocal space possess the same number of sides with identical weight,  (ii) two thermodynamic equilibrium conditions - the stability of the crystal phase and the equality of free energy in coexisting crystal and melt phases\cite{Shih87,Wu07}. These coefficients are $a^{(3)}$=$2\frac{a^{(2)}}{v_\mathrm{c}}+2\frac{b^{(2)}v_\mathrm{c}}{u^2_\mathrm{c}}$, $c_{ijk}$=1/12, $a^{(4)}$=$\frac{b^{(2)}v^2_\mathrm{c}}{u^4_\mathrm{c}}$, $c_{ijkl}$=1/12, $b^{(4)}$=$\frac{a^{(2)}}{v^2_\mathrm{c}}$ and $d_{ijkl}$=1/24, respectively.

{\bf{[Step2]}}: In the limit of low undercooling, the thermodynamic driving force for a CMI outside of equilibrium has the form of $L\frac{\Delta T}{T_\mathrm{m}}$, where $L$ is the latent heat of fusion per particle. Therefore the total free energy of a non-equilibrium fcc/melt CMI system near $T_\mathrm{m}$ has the following form, 
\begin{equation}
 \tag{S4}
\begin{aligned}
&\Delta F_\mathrm{ne}=\Delta F+n_{0}\int \mathrm{d}\vec{r}\\
&\left(\sum_i\frac{c_u}{8}\frac{u_\mathrm{c}-u_i}{u_\mathrm{c}-u_\mathrm{m}}+\sum_i\frac{1-c_u}{6}\frac{v_\mathrm{c}-v_i}{v_\mathrm{c}-v_\mathrm{m}}\right)\frac{\Delta T}{T_\mathrm{m}}L,
\label{eq:fg}
\end{aligned}
\end{equation}
under the assumption that the thermodynamic driving force decays proportionally to the combination of the two-mode density wave amplitudes across the fcc/melt CMIs. We employ here a fraction factor $0<c_u<1$ to ensure the normalization, and linear interpolation functions as in Ref.\onlinecite{Wu15} to model the driving force variation in each contributing terms. We shall see in the following derivations, that the crystal growth velocity is independent of both the value of $c_u$ and the form of the interpolation function.

{\bf{[Step3]}}: According to the well-known standard time-dependent Ginzburg-Landau\cite{Gunton83,Harrowell86} model, the dynamical evolution of the order parameter profiles in the fcc/melt CMI system, $u_i(\vec{r},t)$ and $v_i(\vec{r},t)$, are governed by the equations,
\begin{equation}
 \tag{S5}
\begin{aligned}
\varsigma_{1i}\frac{\partial u_i}{\partial t}&=-\frac{1}{n_0 k_{\mathrm{B}} T} \frac{\delta \Delta F_\mathrm{ne}}{\delta u_i}\\
\varsigma_{2i}\frac{\partial v_i}{\partial t}&=-\frac{1}{n_0 k_{\mathrm{B}} T} \frac{\delta \Delta F_\mathrm{ne}}{\delta v_i}
\end{aligned}
\label{eq:tdgl1}
\end{equation}
the DTCs, $\varsigma_{1i}$=$\tau(|\vec{K}_i|)/S(|\vec{K}_i|)$ and $\varsigma_{2i}$=$\tau(|\vec{G}_i|)/S(|\vec{G}_i|)$, are related to the ratio between the dynamic time scales of the density waves relax at the given RLVs in the melt phases and their corresponding static structure factors. The relaxation time $\tau$ is measured as the inverse half-width of the dynamical structure factor at a specified wavenumber, i.e., $S(|\vec{K}_i|,\omega)$ and $S(|\vec{G}_i|,\omega)$, see in Methods for details.

Within the steady-state crystal growth regime, for a given growth direction along one of the crystal surface normal $\hat{n}$. If one employs the coordinate transformation $z=\hat{n}\cdot\vec{r}-V_\mathrm{I} t$, the order parameter profiles can be transformed into the time-independent form, i.e., $u_i(z)$ and $v_i(z)$. Because $\frac{\partial u_i}{\partial t}$=$\frac{\mathrm{d}u_i}{\mathrm{d}z}\frac{\mathrm{d}z}{\mathrm{d}t}$=$-V_\mathrm{I}\frac{\mathrm{d}u_i}{\mathrm{d}z}$, and $\frac{\partial v_i}{\partial t}$=$\frac{\mathrm{d}v_i}{\mathrm{d}z}\frac{\mathrm{d}z}{\mathrm{d}t}$=$-V_\mathrm{I}\frac{\mathrm{d}v_i}{\mathrm{d}z}$, Eq.\ref{eq:tdgl1} can be written in a form that dictates the CMI within a dynamic frame migrating with constant velocity $V_\mathrm{I}$,
\begin{equation}
 \tag{S6}
\begin{aligned}
V_\mathrm{I}\varsigma_{1i}\frac{\mathrm{d} u_i}{\mathrm{d} z}&=\frac{1}{n_0 k_{\mathrm{B}} T} \frac{\delta \Delta F_\mathrm{ne}}{\delta u_i}\\
V_\mathrm{I}\varsigma_{2i}\frac{\mathrm{d} v_i}{\mathrm{d} z}&=\frac{1}{n_0 k_{\mathrm{B}} T} \frac{\delta \Delta F_\mathrm{ne}}{\delta v_i}
\end{aligned}
\label{eq:tdgl2}
\end{equation}

%~~~~~~~~~~~~~~~ 
%\begin{sidewaystable}[h]
\begin{table*}
\renewcommand{\tablename}{Tab.S}
\caption{List of symbols representing classifications of the density wave amplitude order parameters, the RLV subsets with respect to the fcc/melt CMI normals, and the corresponding dissipative time constants used in TDGL calculation.}
\begin{ruledtabular}
\begin{tabular}{lccccccc}
&\multicolumn{3}{c}{fcc(100)/melt}&\multicolumn{4}{c}{fcc(011)/melt}\\
\hline
Order parameter                                     
&$u$&$v_a$&$v_b$&$u_a$&$u_b$&$v_a$&$v_b$\\
RLV subset                                     
&$\left\langle111\right\rangle$&$\left\langle200\right\rangle$&$\left\langle200\right\rangle$&$\left\langle111\right\rangle$&$\left\langle111\right\rangle$&$\left\langle200\right\rangle$&$\left\langle200\right\rangle$\\
Number of $\vec{K}_i$ or $\vec{G}_i$ 
&8&4&2&4&4&4&2\\
$(\hat{K}_i\cdot\hat{n})^2$ or $(\hat{G}_i\cdot\hat{n})^2$
&1/3&0&1&2/3&0&1/2&0\\
DTC&$\varsigma_1$
&$\varsigma_2$&$\varsigma_2$&$\varsigma_1$&$\varsigma_1$&$\varsigma_2$&$\varsigma_2$
\end{tabular}
\end{ruledtabular}
\label{tabKnGn}
\end{table*} % double column
%\end{sidewaystable}      
%~~~~~~~~~~~~~~~ 

{\bf{[Step4]}}: In this final step, we derive the analytic expression for fcc/melt CMI $\mu$ from the steady-state propagating solution of the TDGL equation set (Eq.\ref{eq:tdgl2}) following the recipe of Wu and Karma\cite{Wu15} for the bcc/melt CMI system.

Because eight principal and six second-set RLVs have different relative orientations with respect to the CMI normal $\hat{n}$, the amplitudes of the density waves decay with different rates and contribute differently to the dissipation of during the crystal growth in different orientations. Tab.S\ref{tabKnGn} summarizes the two fcc/melt CMI orientations, which are fcc(100)/melt and fcc(011)/melt investigated in the current study, with the corresponding categories of $(\hat{K}_i\cdot\hat{n})^2$ and $(\hat{G}_i\cdot\hat{n})^2$. Note that $(\hat{K}_i\cdot\hat{n})^2$ and $(\hat{G}_i\cdot\hat{n})^2$ determines the coefficients of the S-G terms in $\Delta F_\mathrm{ne}$. In Tab.S\ref{tabKnGn} and the following derivations, we use a unified order parameter $u$ to describe a first category of density wave amplitudes because all 8 principal RLVs have the same symmetry with respect to the fcc(100)/melt CMI normal and an identical magnitude of $(\hat{K}_i\cdot\hat{n})^2$. The second set of RLVs is divided into two subsets with different magnitude of $(\hat{G}_i\cdot\hat{n})^2$, and we use order parameters $v_a$ and $v_b$ to separately describe the two subsets of the density wave amplitudes. Correspondingly, the DTCs $\varsigma_{1i}$ and $\varsigma_{2i}$ are summarized with two unified forms ($\varsigma_1=\tau(|\vec{K}_{\langle111\rangle}|)/S(|\vec{K}_{\langle111\rangle}|)$ and $\varsigma_2=\tau(|\vec{G}_{\langle200\rangle}|)/S(|\vec{G}_{\langle200\rangle}|)$) in the Tab.S\ref{tabKnGn}.

With the application of the functional derivatives $\frac{\delta\Delta F}{\delta u_i}$=$n_0 k_{\mathrm{B}}T(\frac{\partial f}{\partial u_i}-\frac{\mathrm{d}}{\mathrm{d}z}\frac{\partial f}{\partial u^{\prime}_i})$, $\frac{\delta\Delta F}{\delta v_i}$=$n_0 k_{\mathrm{B}} T(\frac{\partial f}{\partial v_i}-\frac{\mathrm{d}}{\mathrm{d}z}\frac{\partial f}{\partial v^{\prime}_i})$, the Eq.\ref{eq:tdgl2} transforms to the following form using the unified order parameter categories listed in Tab.S\ref{tabKnGn}, 
\begin{equation}
 \tag{S7}
\begin{aligned}
8V_\mathrm{I}\varsigma_1 \frac{\mathrm{d} u}{\mathrm{d} z}&=f_{u}+4D+8\alpha_{1}\\
4V_\mathrm{I}\varsigma_2 \frac{\mathrm{d} v_a}{\mathrm{d} z}&=f_{v_a}+2E^{a}+4\alpha_{2}\\
2V_\mathrm{I}\varsigma_2 \frac{\mathrm{d} v_b}{\mathrm{d} z}&=f_{v_b}+E^{b}+2\alpha_{2}.
\end{aligned}
\label{eq:tdgl3}
\end{equation}

Here, we choose to employ fcc(100) orientation as an illustration to fulfill the derivation of the analytical expression of $\mu$ for fcc/melt CMIs. In the Eq.\ref{eq:tdgl3}, $f_{u}$, $f_{v_a}$ and $f_{v_b}$ denote the partial derivatives of excess free energy density $f$ with respect to the three categories of order parameters. The rest parameters, such as $D$, $E^{a,b}$ and $\alpha_{1,2}$, are defined as $D$=$C^{\prime\prime}(|\vec{K}_{\langle111\rangle}|)(\hat{K}_{u}\cdot\hat{n})^2\frac{\mathrm{d}^2u}{\mathrm{d}z^2}$, $E^{a,b}$= $\left[C^{\prime\prime}(|\vec{G}_{\langle200\rangle}|)(\hat{G}_{v_{a,b}}\cdot\hat{n})^2+\frac{C^{\prime}(|\vec{G}_{\langle200\rangle}|)}{|\vec{G}_{\langle200\rangle}|}(1-\hat{G}_{v_{a,b}}\cdot\hat{n})^2)\right]$ $\frac{\mathrm{d}^2v_{a,b}}{\mathrm{d}z^2}$, $\alpha_1=\frac{c_uL\Delta T}{8(u_\mathrm{c}-u_\mathrm{m})k_\mathrm{B}T^2_\mathrm{m}}$ and $\alpha_2=\frac{(1-c_u)L\Delta T}{6(v_\mathrm{c}-v_\mathrm{m})k_\mathrm{B}T^2_\mathrm{m}}$, respectively. 

To find solutions, we employ linear perturbation expansions of the first two terms on the right sides of the Eq.\ref{eq:tdgl3} around equilibrium order parameter profiles. At low undercooling limit, order parameters can be viewed as the sum of the stationary equilibrium value at $T_\mathrm{m}$ and small linear perturbations outside of equilibrium, e.g., $u(z)\approx u^{(0)}(z)$+$u^{(1)}(z)$, $v_{a,b}(z)\approx v_{a,b}^{(0)}(z)$+$v_{a,b}^{(1)}(z)$. By further applying boundary conditions of $T=T_\mathrm{m}$ and $V_\mathrm{I}$=$\alpha_1$=$\alpha_2$=0, the Eq.\ref{eq:tdgl3} is linearized into a coupled linear equation set as follows,
\begin{widetext}
\begin{equation}
 \tag{S8}
\begin{aligned}
8V_\mathrm{I}\varsigma_1 \frac{\mathrm{d} u^{(0)}}{\mathrm{d} z}-8\alpha_1&=f_{uu}u^{(1)}+f_{uv_a}v_a^{(1)}+f_{uv_b}v_b^{(1)}+4D_uu^{(1)}\\
4V_\mathrm{I}\varsigma_2 \frac{\mathrm{d} v_a^{(0)}}{\mathrm{d} z}-4\alpha_2&=f_{v_au}u^{(1)}+f_{v_av_a}v_a^{(1)}+f_{v_av_b}v_b^{(1)}+2E^{a}_{v_a}v_a^{(1)}\\
2V_\mathrm{I}\varsigma_2 \frac{\mathrm{d} v_b^{(0)}}{\mathrm{d} z}-2\alpha_2&=f_{v_bu}u^{(1)}+f_{v_bv_a}v_a^{(1)}+f_{v_bv_b}v_b^{(1)}+E^{b}_{v_b}v_b^{(1)},
\end{aligned}
\label{eq:tdgl4}
\end{equation}
or alternatively, representing with matrix notations, {\bf AX}={\bf B},
\begin{equation}
 \tag{S9}
\begin{aligned}
\mathrm{\bf A}=\left(\begin{array}{c}{u^{(1)}} \\ {v_a^{(1)}} \\ {v_b^{(1)}}\end{array}\right),
\mathrm{\bf X}=\left(\begin{array}{ccc}
{f_{uu}+4D_u}&{f_{uv_a}}                     &{f_{uv_b}}\\
{f_{v_au}}     &{f_{v_av_a}+2E^{a}_{v_a}} &{f_{v_av_b}}\\
{f_{v_bu}}     &{f_{v_bv_a}}                  &{f_{v_bv_b}+E^{b}_{v_b}}
\end{array}\right),
\mathrm{\bf B}=\left(\begin{array}{l}
{8V_\mathrm{I}\varsigma_{1}\frac{\mathrm{d} u^{(0)}}{\mathrm{d} z}-8\alpha_{1}}\\
{4V_\mathrm{I}\varsigma_{2}\frac{\mathrm{d} v_a^{(0)}}{\mathrm{d} z}-4\alpha_{2}}\\
{2V_\mathrm{I}\varsigma_{2}\frac{\mathrm{d} v_b^{(0)}}{\mathrm{d} z}-2\alpha_{2}}
\end{array}\right)
\label{eq:tdgl5}
\end{aligned}
\end{equation}

If one links two distinct features of the linear operators with the solvability condition of the above matrix equation, the analytical expression for the kinetic coefficient $\mu$ can thus be obtained. The first feature is ${\bf A}^{(0)}{\bf X}$=0 , where column vector function ${\bf A}^{(0)}$ has three components $\frac{\mathrm{d} u^{(0)}}{\mathrm{d} z}$, $\frac{\mathrm{d} v_a^{(0)}}{\mathrm{d} z}$ and $\frac{\mathrm{d} v_b^{(0)}}{\mathrm{d} z}$\cite{Wu15}. The second features is that the symmetric matrix ${\bf X}$ is a self-adjoint operator, so that the inner products of the row vectors ${\bf A}^{(0){\bf T}}$ (or ${\bf A}^{\bf T}$) and the column vectors ${\bf AX}$ (or${\bf A}^{(0)}{\bf X}$) are equal, i.e., (${\bf A}^{(0){\bf T}}$,${\bf AX}$)=(${\bf A}^{\bf T}$,${\bf A}^{(0)}{\bf X}$) where ${\bf A}^{(0){\bf T}}$ is the transposed row vector function of ${\bf A}^{(0)}$. The two features naturally draw out the solvability condition of (${\bf A}^{(0){\bf T}},{\bf B}$)=$\int_{-\infty}^{+\infty} \mathrm{d} z {\bf A}^{{(0)}{{\bf T}}}\cdot{\bf B}$=0, thus yeilding,
\begin{equation}
 \tag{S10}
\begin{aligned}
\int_{-\infty}^{+\infty} \mathrm{d} z \left\{V_\mathrm{I}
\left[8\varsigma_1(\frac{\mathrm{d}u^{(0)}}{\mathrm{d}z})^2+
4\varsigma_2(\frac{\mathrm{d}v_a^{(0)}}{\mathrm{d}z})^2+
2\varsigma_2(\frac{\mathrm{d}v_b^{(0)}}{\mathrm{d}z})^2\right]-
\left[8\alpha_1\frac{\mathrm{d}u^{(0)}}{\mathrm{d}z}+4\alpha_2\frac{\mathrm{d}v_a^{(0)}}{\mathrm{d}z}+2\alpha_2\frac{\mathrm{d}v_b^{(0)}}{\mathrm{d}z}\right]\right\}=0.
\label{eq:tdgl6}
\end{aligned}
\end{equation}
\end{widetext}
With given boundary conditions that $u^{(0)}=u_\mathrm{m}$, $v_a^{(0)}=v_b^{(0)}=v_\mathrm{m}$ at $z=-\infty$ and $u^{(0)}=u_\mathrm{c}$, $v_a^{(0)}=v_b^{(0)}=v_\mathrm{c}$ at $z=\infty$, respectively, the above equation can be  further simplified. Besides, to make the following expression to adapt different orientations, we use $u_i$ and $v_i$ of the equilibrium fcc/melt CMIs to replace $u^{(0)}$ and $v_{a,b}^{(0)}$ which are specifically describing fcc(100)/melt CMI in the Eq.\ref{eq:tdgl6},
\begin{equation}
 \tag{S11}
\begin{aligned}
V_\mathrm{I}&=\frac{L\Delta T}{k_\mathrm{B}T^2_\mathrm{m}}\left[\int\mathrm{d}z\varsigma_1\sum_{\vec{K}_i,u_{a,b}}(\frac{\mathrm{d}u_i}{\mathrm{d}z})^2+\varsigma_2\sum_{\vec{G}_i,v_{a,b}}(\frac{\mathrm{d}v_i}{\mathrm{d}z})^2 \right]^{-1}.
\label{eq:tdgl7}
\end{aligned}
\end{equation}
Dividing both sides of Eq.\ref{eq:tdgl7} by $\Delta T$, the analytical expression of the kinetic coefficient for the fcc/melt CMIs is finally obtained,
\begin{equation}
 \tag{S12}
\mu_{\hat{n}}=\frac{L}{k_\mathrm{B}T^2_\mathrm{m}A_{\hat{n}}},
\label{eq:tdgl8}
\end{equation}
in which, the anisotropy factor $A_{\hat{n}}$ for the given CMI orientation $\hat{n}$ is defined as
\begin{equation}
 \tag{S13}
A_{\hat{n}}=\int\mathrm{d}z\Big[\varsigma_1\sum_{\vec{K}_i,u_{a,b}}(\frac{\mathrm{d}u_i}{\mathrm{d}z})^2+\varsigma_2\sum_{\vec{G}_i,v_{a,b}}(\frac{\mathrm{d}v_i}{\mathrm{d}z})^2\Big].
\label{eq:tdgl9}
\end{equation}
The density wave amplitude sets and relating S-G terms listed in Tab.S\ref{tabKnGn} would result in different values of $A_{\hat{n}}$ in predicting $\mu_{\hat{n}}$ for different fcc/melt CMI orientation or growth directions.

We should note here, above TDGL formalism for predicting the CMI kinetic coefficient can be readily extended to more simple elemental crystal structures, such as hcp, hexagonal, etc.

%==============================================================
\vspace{.5cm}
\noindent B. {\it o}bco/({\it o})melt CMIs
\label{ssec:ap-theory-bco}
\vspace{.3cm}
%==============================================================
%==============================================================

%~~~~~~~~~~~~~~~ 
%\begin{sidewaystable}[h]
\begin{table*}%[H]
\renewcommand{\tablename}{Tab.S}
\caption{List of symbols representing classifications of the density wave amplitude order parameters, the RLV subsets with respect to the {\it o}bco/melt and {\it o}bco/{\it o}melt CMI normals, and the corresponding dissipative time constants used in TDGL calculation. The RLV subsets are defined as 
$\vec{K}_{u_a}=(\pm\frac{2\pi}{\mathrm{a_1}},0,\pm\frac{2\pi}{\mathrm{a_3}})$, 
$\vec{K}_{u_b}=(\pm\frac{2\pi}{\mathrm{a_1}}, \pm\frac{2\pi}{\mathrm{a_2}},0)$, 
$\vec{G}_{v_a}=(0,\pm\frac{2\pi}{\mathrm{a_2}},\pm\frac{2\pi}{\mathrm{a_3}})$ and 
$\vec{G}_{v_b}=(\pm\frac{4\pi}{\mathrm{a_1}},0,0)$, respectively.}
\begin{ruledtabular}
\begin{tabular}{lcccccccc}
&\multicolumn{4}{c}{{\it o}bco(100)/({\it o})melt}&\multicolumn{4}{c}{{\it o}bco(001)/({\it o})melt}\\
\hline
Order parameter                                     
&$u_a$&$u_b$&$v_a$&$v_b$&$u_a$&$u_b$&$v_a$&$v_b$\\
RLV subset
&$\vec{K}_{u_a}$
&$\vec{K}_{u_b}$
&$\vec{G}_{v_a}$
&$\vec{G}_{v_b}$
&$\vec{K}_{u_a}$
&$\vec{K}_{u_b}$
&$\vec{G}_{v_a}$
&$\vec{G}_{v_b}$\\
Number of $\vec{K}_i$ or $\vec{G}_i$ 
&4&4&4&2&4&4&4&2\\
$(\hat{K}_i\cdot\hat{n})^2$ or $(\hat{G}_i\cdot\hat{n})^2$
&$\frac{\mathrm{a^2_3}}{\mathrm{a^2_1}+\mathrm{a^2_3}}$
&$\frac{\mathrm{a^2_2}}{\mathrm{a^2_1}+\mathrm{a^2_2}}$
&0
&1
&$\frac{\mathrm{a^2_1}}{\mathrm{a^2_1}+\mathrm{a^2_3}}$
&0
&$\frac{\mathrm{a^2_2}}{\mathrm{a^2_2}+\mathrm{a^2_3}}$
&0\\
DTC&$\varsigma_{1a}$&$\varsigma_{1b}$
&$\varsigma_{2a}$&$\varsigma_{2b}$&$\varsigma_{1a}$&$\varsigma_{1b}$&$\varsigma_{2a}$&$\varsigma_{2b}$
\end{tabular}
\end{ruledtabular}
\label{tabKnGn-bco}
\end{table*} 
%\end{sidewaystable}   
%~~~~~~~~~~~~~~~ 

In what follows, we first introduce the RLVs in the {\it o}bco crystal, then we present the brief derivation of the analytical expression of $\mu$ for the {\it o}bco/melt and {\it o}bco/{\it o}melt CMIs as was shown in the previous subsection for the fcc/melt CMIs.

As mentioned in the Methods and the following Section \ref{sec:ap-EDM} of the Supplemental Information, the {\it o}bco crystal structure originates from fcc crystal with the spontaneous polarization along $[01\bar{1}]$ due to the increasing dipolar interactions. Actually, the {\it o}bco structures near their $T_\mathrm{m}^{*}$ for $M^*$=$\sqrt{2}$ or 2, resemble fcc structures more closely in RLVs than they resemble body centered crystal structures. So we need to keep the second set of RLVs as in the fcc/melt CMIs system. Based on the unit vectors ($\hat{x}$, $\hat{y}$, $\hat{z}$) shown in the Fig.S\ref{fig1}, we choose a set of translation vectors to described the {\it o}bco crystal structure, i.e., $(0,0,\mathrm{a_3})$, $(\frac{\mathrm{a_1}}{2},-\frac{\mathrm{a_2}}{2},\frac{\mathrm{a_3}}{2})$ and $(\frac{\mathrm{a_1}}{2},\frac{\mathrm{a_2}}{2},\frac{\mathrm{a_3}}{2})$. This set of lattice vectors reduces to the primitive vector of the fcc crystal under zero polarization along fcc $[01\bar{1}]$. Accordingly, the corresponding eight principal RLVs $\vec{K}_i$ are $(\pm\frac{2\pi}{\mathrm{a_1}},0,\pm\frac{2\pi}{\mathrm{a_3}})$ and ($\pm\frac{2\pi}{\mathrm{a_1}}, \pm\frac{2\pi}{\mathrm{a_2}},0$), the six second-set RLVs $\vec{G}_i$ are $(0,\pm\frac{2\pi}{\mathrm{a_2}},\pm\frac{2\pi}{\mathrm{a_3}})$ and ($\pm\frac{4\pi}{\mathrm{a_1}},0,0$). Tab.S\ref{tabKnGn-bco} list the categorization of the RLVs, unified order parameters (density wave amplitudes) and and their corresponding contribution in the S-G dissipation terms, for the two orientations of the {\it o}bco/({\it o})melt CMIs. One can find that the magnitude of $(\hat{K}_i\cdot\hat{n})^2$ or $(\hat{G}_i\cdot\hat{n})^2$ reduce to the same value in Tab.S\ref{tabKnGn} in the limit of {\it o}bco reduces to fcc.

We next present, in a concise manner, our formalism of TDGL theory of $\mu$ for {\it o}bco/({\it o})melt CMIs. The derivation process is also consist of four steps.

{\bf{[Step1]}}: For the equilibrium {\it o}bco/({\it o})melt CMIs, due to the presence of the orientational ordering (or polarization) in crystal (or in both crystal and melt), additional terms must be added into the excess free energy $\Delta F$. In Eq.\ref{eq:2f}, $\Delta F$ is expressed as the sum of a term of the $u_i$ and $v_i$ defined in Eq.\ref{eq:denswave} and a term of orientational order parameter ($o$).
\begin{equation}
 \tag{S14}
\Delta F\equiv n_{0}k_\mathrm{B}T\int f(u_i,v_i)\mathrm{d}\vec{r}+n_{0}k_\mathrm{B}|T_\mathrm{C}-T|\int g(o)\mathrm{d}\vec{r}.
\label{eq:2f}
\end{equation}
$u_i$ and $v_i$ take the values of $u_\mathrm{c}$ and $v_\mathrm{c}$ in the {\it o}bco crystals, respectively, and values of 0 in both melt phases. The orientational order parameter $o$ takes the value of $P_\mathrm{c}$ and $P_\mathrm{m}$ in the {\it o}bco crystal and the ({\it o})melt phases, respectively. Note that, $P_\mathrm{m}$ is greater than zero in the ferroelectric {\it o}melt phase. $T_\mathrm{C}$ is the Curie temperature.

The excess free-energy density due to density waves $f(u_i,v_i)$ has the same expression as Eq.\ref{eq:fpower} for the fcc/melt CMIs systems. All the multiplicative coefficients in the $f(u_i,v_i)$ are analogous to the case of Eq.\ref{eq:fpower} while referring to the categorization of RLVs and order parameters in Tab.S\ref{tabKnGn-bco}. The excess free-energy density due to polarization $g(o)$ has the following form,

%\begin{widetext}
\begin{equation}
 \tag{S15}
\begin{aligned} 
&g(o)\approx\frac{1}{2}\kappa^{(2)}o^2+\frac{1}{2}\gamma^{(2)}(\frac{\mathrm{d}o}{\mathrm{d}z})^{2},
\end{aligned}
\label{eq:2fpower}
\end{equation}
%\end{widetext}
here we adopted the lowest order of a ferroelectric Landau-Ginzburg free energy for a ferroelectric system as Chandra and Littlewood assumed\cite{Chandra07}. The two terms in Eq.\ref{eq:2fpower} represent the leading contributions in the bulk polarization free energies, and the free energy arises from the small spatial variation of the polarization, respectively. Note that, we assume here that the density waves and the polarization do not have a strong coupling effect. Evidence supports this assumption is found and provided in Fig.S\ref{figS2}. The coefficients in Eq.\ref{eq:2fpower} are defined as $\kappa^{(2)}=\frac{\kappa}{n_0k_\mathrm{B}|T_\mathrm{C}-T|}$ and $\gamma^{(2)}=\frac{\gamma}{n_0k_\mathrm{B}|T_\mathrm{C}-T|}$. $\kappa$ is the dielectric stiffness (proportional to $|T_\mathrm{C}-T|$), and $\gamma$ is the per unit volume free-energy cost of forming parallel polarization.

\begin{widetext}
{\bf{[Step2]}}: The thermodynamic driving force for the non-equilibrium {\it o}bco/({\it o})melt CMI is assumed to decay proportionally to both the two-mode density wave amplitudes and the Curie temperature-dependent polarization, 
\begin{equation}
 \tag{S16}
\begin{aligned}
&\Delta F_\mathrm{ne}=\Delta F\\
&+n_{0}k_{\mathrm{B}} T\int \mathrm{d}\vec{r}\left(\sum_i\frac{c_u}{8}\frac{u_\mathrm{c}-u_i}{u_\mathrm{c}-u_\mathrm{m}}+\sum_i\frac{c_v}{6}\frac{v_\mathrm{c}-v_i}{v_\mathrm{c}-v_\mathrm{m}}\right)\frac{\Delta T}{T_\mathrm{m}}\frac{L}{k_{\mathrm{B}} T}+n_{0}k_{\mathrm{B}} |T_\mathrm{C}-T|\int \mathrm{d}\vec{r}\left(c_o\frac{P_\mathrm{c}-o}{P_\mathrm{c}-P_\mathrm{m}}\right)\frac{\Delta T}{T_\mathrm{m}}\frac{L}{k_{\mathrm{B}} T},
\label{eq:2fg}
\end{aligned}
\end{equation}
where the fraction factors ensures the sum of variational contributions are normalized to 1, i.e., $c_u+c_v+c_o$ =1.
\end{widetext}

{\bf{[Step3]}}: One standard ferroelectric TDGL dynamic equation is added to describe the dynamical evolution of the polarization order parameter profiles, $o(\vec{r},t)$, which migrates together with $u_i(\vec{r},t)$ and $v_i(\vec{r},t)$. The validity of the simultaneous migration assumed here is verified below in Fig.S\ref{figS1}.
\begin{equation}
 \tag{S17}
\begin{aligned}
V_\mathrm{I}\varsigma_{1i}\frac{\mathrm{d} u_i}{\mathrm{d} z}&=\frac{1}{n_0 k_{\mathrm{B}} T} \frac{\delta \Delta F_\mathrm{ne}}{\delta u_i}\\
V_\mathrm{I}\varsigma_{2i}\frac{\mathrm{d} v_i}{\mathrm{d} z}&=\frac{1}{n_0 k_{\mathrm{B}} T} \frac{\delta \Delta F_\mathrm{ne}}{\delta v_i}\\
V_\mathrm{I}\varsigma_3\frac{\mathrm{d} o}{\mathrm{d} z}  &=\varepsilon_0\chi\frac{\delta \Delta F_\mathrm{ne}}{\delta o}.
\end{aligned}
\label{eq:2tdgl2}
\end{equation}
Here, the polarization along with the crystallization is assumed to respond linearly with the electric-field across the CMIs. The functional derivatives of $\Delta F_\mathrm{ne}$ with respect of polarization density has the dimension of the electric field, $\varepsilon_0$ is the vacuum dielectric permittivity and the susceptibility $\chi=(\varepsilon_\mathrm{r}-1)$ is related to its relative dielectric permittivity $\varepsilon_{\textrm {r}}$. An additional DTC, $\varsigma_3$=$\tau_{o}(q_1^{\mathrm{P}})/S_{o}(q_1^{\mathrm{P}})$, is introduced together with the dynamic equation. $S_{o}(q_1^{\mathrm{P}})$ is the first peak value of the generalized static structure response (dipole-dipole spatial correlation) function, here the subscript ``$o$'' stands for the orientational order parameter related property, and the superscript ``P'' on the wavenumber $q$ stands for the component parallel to the polarization direction. $\tau_{o}(q_1^{\mathrm{P}})$ is the orientational (or dipole rotational) relaxation time which is measured as the inverse half-width of the generalized dynamical dipolar structure response function ($S_{o}(q_1^{\mathrm{P}},\omega)$) of the melt phase, see in Methods for details.

{\bf{[Step4]}}: Based on the four categories of the unified order parameter in Tab.S\ref{tabKnGn-bco}, the Eq.\ref{eq:2tdgl2} transforms to a system with five linear equations, describing the cooperatively dissipation the excess free energy of the non-equilibrium ($\Delta F_\mathrm{ne}$) {\it o}bco/({\it o})melt CMIs,
\begin{equation}
 \tag{S18}
\begin{aligned}
4V_\mathrm{I}\varsigma_{1a}\frac{\mathrm{d} u_a}{\mathrm{d} z}&=f_{u_a}+2D^{a}+4\alpha_{1}\\
4V_\mathrm{I}\varsigma_{1b}\frac{\mathrm{d} u_b}{\mathrm{d} z}&=f_{u_b}+2D^{b}+4\alpha_{1}\\
4V_\mathrm{I}\varsigma_{2a}\frac{\mathrm{d} v_a}{\mathrm{d} z}&=f_{v_a}+2E^{a}+4\alpha_{2}\\
2V_\mathrm{I}\varsigma_{2b}\frac{\mathrm{d} v_b}{\mathrm{d} z}&=f_{v_b}+E^{b}+2\alpha_{2}\\
V_\mathrm{I}\varsigma_3\Omega\frac{\mathrm{d} o}{\mathrm{d} z}&=g_{o}+O+\alpha_{3},
\end{aligned}
\label{eq:2tdgl3}
\end{equation}
%$\varsigma_{1a}=\tau(|\vec{K}_{10\frac{\mathrm{a_1}}{\mathrm{a_3}}}|)/S(|\vec{K}_{10\frac{\mathrm{a_1}}{\mathrm{a_3}}}|)$, $\varsigma_{1b}=\tau(|\vec{K}_{1\frac{\mathrm{a_1}}{\mathrm{a_2}}0}|)/S(|\vec{K}_{1\frac{\mathrm{a_1}}{\mathrm{a_2}}0}|)$, $\varsigma_{2a}=\tau(|\vec{G}_{200}|)/S(|\vec{G}_{200}|)$ and $\varsigma_{2b}=\tau(|\vec{G}_{0\frac{\mathrm{a_1}}{\mathrm{a_2}}\frac{\mathrm{a_1}}{\mathrm{a_3}}}|)/S(|\vec{G}_{0\frac{\mathrm{a_1}}{\mathrm{a_2}}\frac{\mathrm{a_1}}{\mathrm{a_3}}}|)$,
in which, $\Omega=\frac{1}{\varepsilon_0\chi n_0 k_\mathrm{B}|T_\mathrm{C}-T_\mathrm{m}|}$. We use {\it o}bco(100)/melt orientation as an example in the following derivation, and the classifications of the DTCs (in Tab.S\ref{tabKnGn-bco}) with respect to the unified density wave amplitudes order parameters are also employed, $\varsigma_{1a}=\tau(|\vec{K}_{u_a}|)/S(|\vec{K}_{u_a}|)$,$\varsigma_{1b}=\tau(|\vec{K}_{u_b}|)/S(|\vec{K}_{u_b}|)$, $\varsigma_{2a}=\tau(|\vec{G}_{v_a}|)/S(|\vec{G}_{v_a}|)$ and $\varsigma_{2b}=\tau(|\vec{G}_{v_b}|)/S(|\vec{G}_{v_b}|)$. In the above linear equation, $D^{a,b}$, $E^{a,b}$ and $O$ are defined as $D^{a,b}$=$C^{\prime\prime}(|\vec{K}_{u_{a,b}}|)(\hat{K}_{u_{a,b}}\cdot\hat{n})^2\frac{\mathrm{d}^2u_{a,b}}{\mathrm{d}z^2}$, $E^{a,b}$= $\left[C^{\prime\prime}(|\vec{G}_{v_{a,b}}|)(\hat{G}_{v_{a,b}}\cdot\hat{n})^2+\frac{C^{\prime}(|\vec{G}_{v_{a,b}}|)}{|\vec{G}_{v_{a,b}}|}(1-\hat{G}_{v_{a,b}}\cdot\hat{n})^2)\right]$ $\frac{\mathrm{d}^2v_{a,b}}{\mathrm{d}z^2}$ and $O$=$-\frac{\gamma}{n_0k_\mathrm{B}|T_\mathrm{C}-T_\mathrm{m}|}\frac{\mathrm{d}^2o}{\mathrm{d}z^2}$, respectively. Three dimensionless parameters concerning thermodynamic driving force are defined as $\alpha_1=\frac{c_uL\Delta T}{8(u_\mathrm{c}-u_\mathrm{m})k_\mathrm{B}T^2_\mathrm{m}}$,
$\alpha_2=\frac{c_vL\Delta T}{6(v_\mathrm{c}-v_\mathrm{m})k_\mathrm{B}T^2_\mathrm{m}}$, and $\alpha_3=\frac{c_oL\Delta T}{(P_\mathrm{c}-P_\mathrm{m})k_\mathrm{B}T^2_\mathrm{m}}$.

At the low undercooling limit, $u_{a,b}(z)\approx u_{a,b}^{(0)}(z)$+$u_{a,b}^{(1)}(z)$, $v_{a,b}(z)\approx v_{a,b}^{(0)}(z)$+$v_{a,b}^{(1)}(z)$ and $o(z)\approx o^{(0)}(z)$+$o^{(1)}(z)$. Expanding the first two term on the right side of Eq.\ref{eq:2tdgl3} around their equilibrium values, we get
\begin{widetext}
\begin{equation}
 \tag{S19}
\begin{aligned}
4V_\mathrm{I}\varsigma_{1a}\frac{\mathrm{d} u_a^{(0)}}{\mathrm{d} z}-4\alpha_1&=f_{u_a}(u_a^{(0)})+f_{u_au_a}u_a^{(1)}+f_{u_au_b}u_b^{(1)}+f_{u_av_a}v_a^{(1)}+f_{u_av_b}v_b^{(1)}+2D^a(u_a^{(0)})+2D^a_{u_a}u_a^{(1)}\\
4V_\mathrm{I}\varsigma_{1b}\frac{\mathrm{d} u_b^{(0)}}{\mathrm{d} z}-4\alpha_1&=f_{u_b}(u_b^{(0)})+f_{u_bu_a}u_a^{(1)}+f_{u_bu_b}u_b^{(1)}+f_{u_bv_a}v_a^{(1)}+f_{u_bv_b}v_b^{(1)}+2D^b(u_b^{(0)})+2D^b_{u_b}u_b^{(1)}\\
4V_\mathrm{I}\varsigma_{2a}\frac{\mathrm{d} v_a^{(0)}}{\mathrm{d} z}-4\alpha_2&=f_{v_a}(v_a^{(0)})+f_{v_au_a}u_a^{(1)}+f_{v_au_b}u_b^{(1)}+f_{v_av_a}v_a^{(1)}+f_{v_av_b}v_b^{(1)}+2E^a(v_a^{(0)})+2E^a_{v_a}v_a^{(1)}\\
2V_\mathrm{I}\varsigma_{2b}\frac{\mathrm{d} v_b^{(0)}}{\mathrm{d} z}-2\alpha_2&=f_{v_b}(v_b^{(0)})+f_{v_bu_a}u_a^{(1)}+f_{v_bu_b}u_b^{(1)}+f_{v_bv_a}v_a^{(1)}+f_{v_bv_b}v_b^{(1)}+E^b(v_b^{(0)})+E^b_{v_b}v_b^{(1)}\\
V_\mathrm{I}\varsigma_3 \Omega\frac{\mathrm{d} o^{(0)}}{\mathrm{d} z}-\alpha_3&=g_{o}(o^{(0)})+g_{oo}o^{(1)}+O(o^{(0)})+O_{o}o^{(1)}
\end{aligned}
\label{eq:2tdgl4}
\end{equation}
Under the boundary conditions that $T=T_\mathrm{m}$ and $V_\mathrm{I}$=$\alpha_1$=$\alpha_2$=$\alpha_3$=0, the sums of the leading terms in the above expansions equal to zero so that the Eq.\ref{eq:2tdgl4} can be written into matrix equation of the form {\bf AX}={\bf B}. Each matrix notation term is defines as,
\begin{equation}
 \tag{S20}
\begin{aligned}
\mathrm{\bf A}=\left(\begin{array}{c}
{u_a^{(1)}} \\ {u_b^{(1)}} \\ {v_a^{(1)}} \\ {v_b^{(1)}} \\ {o^{(1)}}
\end{array}\right),
\mathrm{\bf X}=\left(\begin{array}{ccccc}
f_{u_au_a}+2D^a_{u_a}&f_{u_au_b}&f_{u_av_a}&f_{u_av_b}&0\\
f_{u_bu_a}&f_{u_bu_b}+2D^b_{u_b}&f_{u_bv_a}&f_{u_bv_b}&0\\
f_{v_au_a}&f_{v_au_b}&f_{v_av_a}+2E^a_{v_a}&f_{v_av_b}&0\\
f_{v_bu_a}&f_{v_bu_b}&f_{v_bv_a}&f_{v_bv_b}+E^b_{v_b}&0\\
0&0&0&0&g_{oo}+O_{o}\\
\end{array}\right),
\mathrm{\bf B}=\left(\begin{array}{c}
{4V_\mathrm{I}\varsigma_{1a}\frac{\mathrm{d} u_a^{(0)}}{\mathrm{d} z}-4\alpha_{1}}\\
{4V_\mathrm{I}\varsigma_{1b}\frac{\mathrm{d} u_b^{(0)}}{\mathrm{d} z}-4\alpha_{1}}\\
{4V_\mathrm{I}\varsigma_{2a}\frac{\mathrm{d} v_a^{(0)}}{\mathrm{d} z}-4\alpha_{2}}\\
{2V_\mathrm{I}\varsigma_{2b}\frac{\mathrm{d} v_b^{(0)}}{\mathrm{d} z}-2\alpha_{2}}\\
{V_\mathrm{I}\varsigma_{3}\Omega\frac{\mathrm{d} o^{(0)}}{\mathrm{d} z}-\alpha_{3}}
\end{array}\right)
\label{eq:2tdgl5}
\end{aligned}
\end{equation}
the solvability condition of this linear problem is obtained similarly as in the previous fcc/melt CMI system, that is 
\begin{equation}
 \tag{S21}
\begin{aligned}
&(\textbf{A}^{(0)\textbf{T}},\textbf{B})=\int_{-\infty}^{+\infty} \mathrm{d} z \Bigg\{V_\mathrm{I}
\left[
4\varsigma_{1a}(\frac{\mathrm{d}u_a^{(0)}}{\mathrm{d}z})^2+
4\varsigma_{1b}(\frac{\mathrm{d}u_b^{(0)}}{\mathrm{d}z})^2+
4\varsigma_{2a}(\frac{\mathrm{d}v_a^{(0)}}{\mathrm{d}z})^2+
2\varsigma_{2b}(\frac{\mathrm{d}v_b^{(0)}}{\mathrm{d}z})^2+
  \varsigma_3 \Omega  (\frac{\mathrm{d}   o^{(0)}}{\mathrm{d}z})^2 \right]\\
&-\left[
4\alpha_1\frac{\mathrm{d}u_a^{(0)}}{\mathrm{d}z}+
4\alpha_1\frac{\mathrm{d}u_b^{(0)}}{\mathrm{d}z}+
4\alpha_2\frac{\mathrm{d}v_a^{(0)}}{\mathrm{d}z}+
2\alpha_2\frac{\mathrm{d}v_b^{(0)}}{\mathrm{d}z}+
\alpha_3\frac{\mathrm{d}o^{(0)}}{\mathrm{d}z}
\right] \Bigg\}=0.
\label{eq:2tdgl6}
\end{aligned}
\end{equation}
Further simplification is made by employing the boundary conditions of order parameters, i.e., $u_{a,b}^{(0)}=u_\mathrm{m}$, $v_{a,b}^{(0)}=v_\mathrm{m}$, $o^{(0)}=P_\mathrm{m}$ at $z=-\infty$ and $u_{a,b}^{(0)}=u_\mathrm{c}$, $v_{a,b}^{(0)}=v_\mathrm{c}$, $o^{(0)}=P_\mathrm{c}$ at $z=\infty$. We then get the analytical expression of the kinetic coefficient $\mu_{100}$ for the given CMI orientation $\hat{n}$ (e.g., {\it o}bco(100)),
\begin{equation}
 \tag{S22}
\mu_{\hat{n}}=\frac{L}{k_\mathrm{B}T^2_\mathrm{m}A_{\hat{n}}},
\label{eq:2tdgl8}
\end{equation}
in which, the anisotropy factor $A_{\hat{n}}$ is defined as
\begin{equation}
 \tag{S23}
A_{\hat{n}}=\int\mathrm{d}z\Big[\varsigma_{1a}\sum_{\vec{K}_i,u_a}(\frac{\mathrm{d}u_i}{\mathrm{d}z})^2+\varsigma_{1b}\sum_{\vec{K}_i,u_b}(\frac{\mathrm{d}u_i}{\mathrm{d}z})^2+\varsigma_{2a}\sum_{\vec{G}_i,v_a}(\frac{\mathrm{d}v_i}{\mathrm{d}z})^2+\varsigma_{2b}\sum_{\vec{G}_i,v_b}(\frac{\mathrm{d}v_i}{\mathrm{d}z})^2+\varsigma_3\Omega(\frac{\mathrm{d}o}{\mathrm{d}z})^2\Big].
\label{eq:2tdgl9}
\end{equation}
The Eq.\ref{eq:2tdgl8} has the identical form as that of fcc/melt CMI system, however, the anisotropy factor $A_{\hat{n}}$ of the {\it o}bco/melt CMIs has more contents. Despite the fact that Eq.\ref{eq:2tdgl8} and Eq.\ref{eq:2tdgl9} are derived for the {\it o}bco(100)/melt CMI orientation, they are certainly applicable for {\it o}bco/{\it o}melt CMIs, and other CMI orientations.
\end{widetext}

\begin{figure}[ht]
\includegraphics[width=3.2 in]{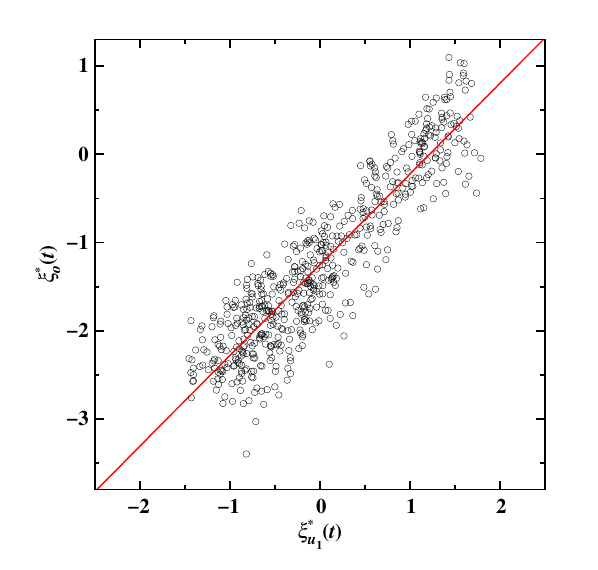}
\renewcommand{\figurename}{Fig.S}
\caption{Linear positive correlation of the two instantaneous interface positions, $\xi^*_{u_1}(t)$ and $\xi^*_{o}(t)$, extracted from the instantaneous density wave amplitude order parameter profile $u_1(z,t)$ and the polarization density profile $o(z,t)$, respectively. Superscript asterisk denote dimensionless length unit. Solid line is a linear fit to the data points.}
\label{figS1}
\end{figure}

%==============================================================
\vspace{.5cm}
\noindent C. Supporting Evidences
\vspace{.3cm}
%==============================================================

To verify the validity of the argument that the dynamical evolution of the polarization order parameter profiles migrates together with the density wave amplitude order parameter profiles. We track both the time evolution of the interface positions $\xi_{u_1}(t)$ and $\xi_{o}(t)$, which are extracted from fits of instantaneous density wave amplitude order parameter profile $u_1(z,t)$ and the orientational order parameter (polarization density) profile $o(z,t)$ to hyperbolic tangent functions at the same MD time step. See in Methods for the more details. Fig.S\ref{figS1} represents a scatter plot of $\xi_{u_1}(t)$ and $\xi_{o}(t)$ for the 500 successive MD trajectories (each separated with 100 MD time steps) of an equilibrium {\it o}bco(100)/melt CMIs, the linear positive correlation of the $\xi_{u_1}(t)$ and $\xi_{o}(t)$ indicate that the orientational order parameter (polarization density) profile migrate simultaneously with the density wave amplitude order parameter profiles.

\begin{figure}[ht]
\includegraphics[width=3.2 in]{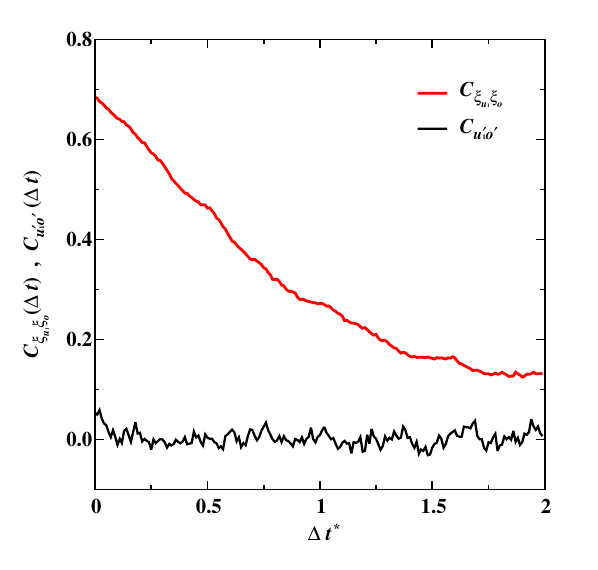}
\renewcommand{\figurename}{Fig.S}
\caption{The cross correlation functions, $C_{u^{\prime}_1o^{\prime}}(\Delta t)$ (solid line) and $C_{\xi_{u_1}\xi_{o}}(\Delta t)$ (dashed line). $\Delta t$ is measured in the reduced unit.}
\label{figS2}
\end{figure}

We provide here an evidence to evaluate one assumption made above (i.e., density waves and the polarization do not have a strong coupling effect and the ferroelectric Landau-Ginzburg free energy terms could be incorporated linearly) by the calculation of the cross correlation function between two time series of instantaneous gradients at interface position of the $u_1(z,t)$ and $o(z,t)$ profiles. The cross correlation function employs the conventional Pearson's cross correlation function form, $C_{XY}(\Delta t)=\mathrm{cov}\{X(t_1),Y(t_2)\}/(\mathrm{var}\{X(t_1)\}\mathrm{var}\{Y(t_2\})^{1/2}$, $\Delta t = |t_1-t_2|$. Here we employ gradient values of the density wave amplitude order parameter profile and the polarization density profile as $X$ and $Y$, specifically, $X(t)=\frac{{\partial}u_1(z,t)}{{\partial}z}|_{z=\xi_{u_1}(t)}$ and $Y(t)=\frac{{\partial}o(z,t)}{{\partial}z}|_{z=\xi_{o}(t)}$, respectively. As a comparison, we also employ $\xi_{u_1}(t)$ and $\xi_{o}(t)$ as $X(t)$ and $Y(t)$ to produce a cross correlation between two interface positions of the two profiles. The results of $C_{u^{\prime}_1o^{\prime}}(\Delta t)$ and $C_{\xi_{u_1}\xi_{o}}(\Delta t)$ are plotted in Fig.S\ref{figS2}. The $C_{\xi_{u_1}\xi_{o}}(\Delta t)$ exhibits an apparent positive correlation with a reasonable decay, consistent with the finding in Fig.S\ref{figS1}. In contrast, the result of $C_{u^{\prime}_1o^{\prime}}(\Delta t)$ suggests that two gradients $\frac{{\partial}u_1(z,t)}{{\partial}z}|_{z=\xi_{u_1}(t)}$ and $\frac{{\partial}o(z,t)}{{\partial}z}|_{z=\xi_{o}(t)}$ are mutually independent, indicating no evident coupling effect is observed between the ordering fluxes regarding polarization and density waves.

%==============================================================
\vspace{.5cm}
\noindent {\bf II. Supporting Information of the Extended Dipole Model}
\label{sec:ap-EDM}
\vspace{.3cm}
%==============================================================

We employ the Extended Dipole Model\cite{Ballenegger04} (EDM) for describing dipolar particles. Two opposite point charges $\pm q$ are fixed symmetrically at $d_q/2$ from the center of each particle, so that each particle carries a permanent dipole moment $M$=$qd_q$, as shown in Fig.S\ref{fig1}. In addition to the Coulombic interaction, the particles also interact with a Lennard-Jones (LJ) potential between the particle centers, $U$=$U_\mathrm{LJ}+U_\mathrm{Coul.}$. We adopt a modified form of LJ pair potential due to Broughton and Gilmer\cite{Broughton83}, 
\begin{equation}
 \tag{S24}
U_\mathrm{LJ}(r)=
\left\{\begin{array}{ll}
{4 \epsilon\left[\left(\frac{\sigma}{r}\right)^{12}-\left(\frac{\sigma}{r}\right)^{6}\right]+C_{1},} & {r \leq 2.3 \sigma} \\ 
{C_{2}\left(\frac{\sigma}{r}\right)^{12}+C_{3}\left(\frac{\sigma}{r}\right)^{6}} \\
{+C_{4}\left(\frac{r}{\sigma}\right)^{2}+C_{5}}, & {2.3 \sigma<r<2.5 \sigma} \\ 
{0.}, & {2.5 \sigma \leq r}
\end{array}\right.
\end{equation}
in which, $C_{1}$=0.016132$\epsilon$, $C_{2}$=3136.6$\epsilon$, $C_{3}$=-68.069$\epsilon$, $C_{4}$=-0.083312$\epsilon$, and $C_{5}$=0.74689$\epsilon$\cite{Davidchack03,Wang13}. This modified form of LJ ensures both the force and the potential energy go to zero smoothly at 2.5$\sigma$. Ballenegger et al. pointed out that EMD behaves similar to the Stockmayer model in describing the dielectric property when $d^*\leqslant0.25$\cite{Ballenegger04}, we shall see below more similarities are found in the structural and thermodynamics properties.

%~~~~~~~~~~~~~~~
\begin{figure}[ht]
\includegraphics[width= 2.6 in]{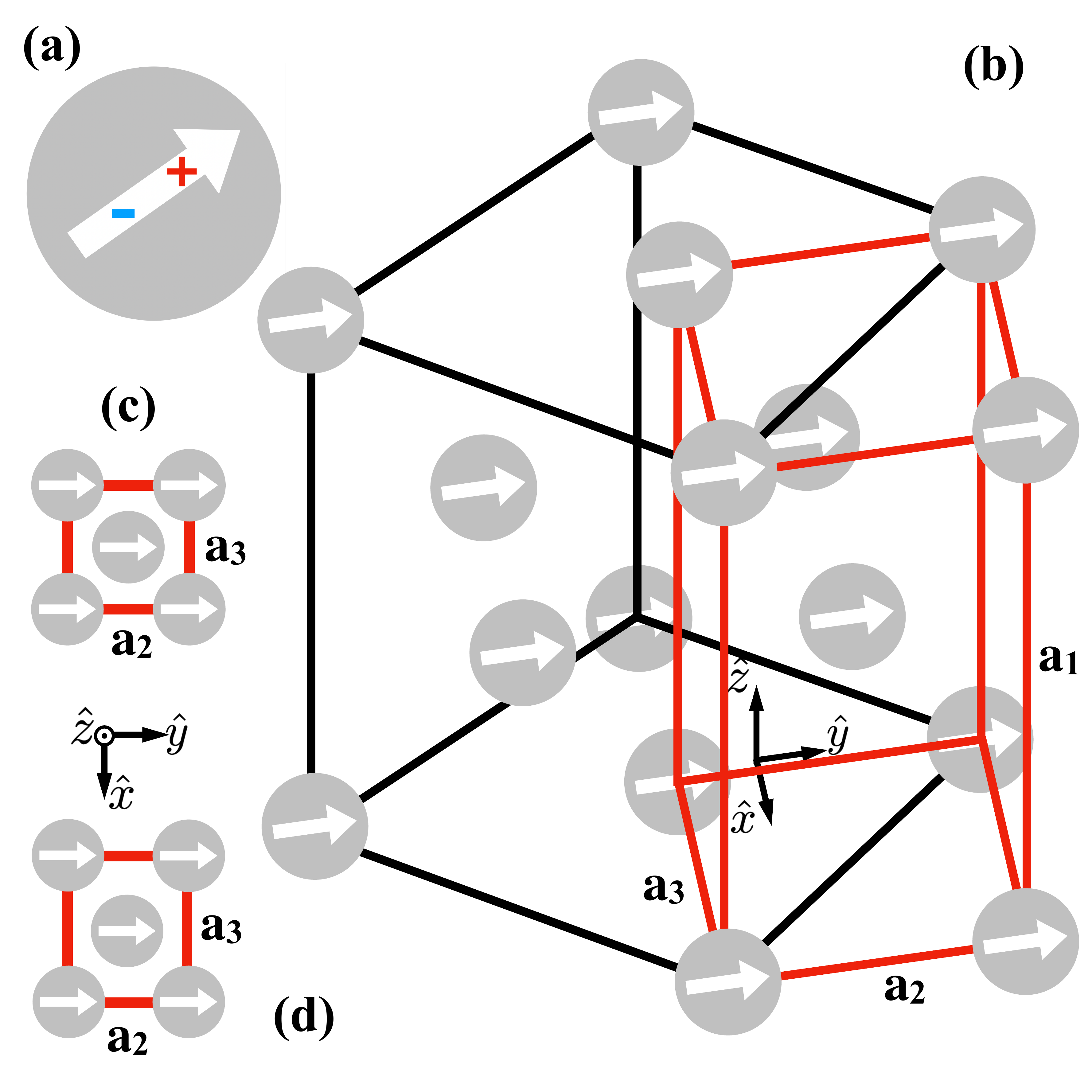}
\renewcommand{\figurename}{Fig.S}
\caption{(a) Schematic diagram of the Extended Dipole Model (EDM) particle, the white arrows show the direction of the particle dipole moment. (b) The orientationally ordered (or ferroelectric) body-centered orthorhombic ({\it o}bco) crystal structure in three dimensions. The {\it o}bco crystal structure reduces to fcc crystal structure under diminishing polarization along unit vector $\hat{y}$. In this limit, [100], [010] and [001] crystallographic orientations of the {\it o}bco structure are analogous to $[100]$, $[01\bar{1}]$ and $[011]$(or $[0\bar{1}\bar{1}]$) of the fcc structure, respectively. (c) Top view of the {\it o}bco structure, particle dipole moment $M^*=\sqrt{2}$, $\mathrm{a_3}/\mathrm{a_2}=1.042$. (d) Top view of the {\it o}bco structure, particle dipole moment $M^*=2$, $\mathrm{a_3}/\mathrm{a_2}=1.135$.}
\label{fig1}
\end{figure}
%~~~~~~~~~~~~~~~

%~~~~~~~~~~~~~~~ 
%\begin{sidewaystable}[h]
\begin{table}[]
\renewcommand{\tablename}{Tab.S}
\caption{Summary of the crystal/melt coexistence properties (number density, axis ratios, and polarization density) for the five $M^*$ systems. Numbers in parentheses are 95\% confidence intervals on the last significant figures.}
\begin{ruledtabular}
\begin{tabular}{ccccccll}
$M^{*}$&$n_{\rm m}^{*}$&$n_{\rm c}^{*}$&a$^*_\mathrm{1}$&a$_\mathrm{1}$/a$_\mathrm{3}$&a$_\mathrm{3}$/a$_\mathrm{2}$&$P_{\rm m}^{*}$&$P_{\rm c}^{*}$\\
\hline
0               &0.830(1)&0.9447(6)&1.618&$\sqrt{2}$&1       &-&-\\
1/2            &0.833(1)&0.9473(4)&1.616&$\sqrt{2}$&1       &0.000(1)&0.002(1)\\
1               &0.857(1)&0.9689(4)&1.604&$\sqrt{2}$&1       &0.001(3)&0.003(4)\\
$\sqrt{2}$  &0.888(1)&1.0095(4)&1.586&1.391       &1.042&0.006(9)&0.984(6)\\
2               &0.920(1)&1.0465(3)&1.597&1.370       &1.135&1.246(3)&1.792(1)\\
\end{tabular}
\end{ruledtabular}
\label{tab1}
\end{table}
%\end{sidewaystable} 
%~~~~~~~~~~~~~~~ 

It has been recognized that orientational order is promoted by the strong and anisotropic\cite{Tao91} dipolar interaction among the dipolar particles, and the system can spontaneously form ferroelectric/ferromagnetic crystal or melt phases for sufficiently large particle dipole moments\cite{Wei92,Klapp98,Groh01,Weis05,Spiteri17,Chen20}. We found in the EDM systems with $M^*$=1, $\sqrt{2}$, 2, that the stable crystal phase (at low temperature) is the {\it o}bco crystal rather than a fcc crystal observed in systems with $M^*$=0, $\frac{1}{2}$, which is in agreement with the stable solid phase predicted for the zero pressure Stcokmayer system\cite{Gao00}. As shown in the Fig.S\ref{fig1}(b), {\it o}bco lattice structure consists of three unit cell lengths $\mathrm{a_1}$, $\mathrm{a_2}$ and $\mathrm{a_3}$, and reduces to fcc lattice with $\mathrm{a_2}$=$\mathrm{a_3}$=$\mathrm{a_1}/\sqrt{2}$ under diminishing polarization along fcc $[01\bar{1}]$. The axis ratio $\mathrm{a_3}/\mathrm{a_2}$ increases with the degree of the bulk polarization as illustrated in Fig.S\ref{fig1}(c) and (d).

As shown in FIG.1 in the main text, we employ $n$=$N_\mathrm{p}/V$ to locate the region in which crystal/melt coexists, and employ polarization density $P$=${\langle 1/V |\sum_{i=1}^{N_\mathrm{p}}\vec{M}_i|\rangle}$ (the summation runs over the $i$th particle of the $N_\mathrm{p}$ dipolar particles in the simulation cell with volume $V$) to monitor the spontaneous polarization or the orientational order. With increasing particle dipole moment, the temperature interval of the $n$ hysteresis increases, associated with a gradual rising melting point $T_\mathrm{m}$ (see in TABLE.I and FIG.2 in the main text). For $M^*$=1 and $\sqrt{2}$, non-zero $P$ at low $T$ corresponds to the {\it o}bco crystal phase (throughout this work, we use dimensionless reduced units which are denoted with superscript asterisks. e.g., reduced temperature $T^*$=$k_\mathrm{B}T/\epsilon$, particle dipole moment $M^*$=$M/\sqrt{\sigma^3\epsilon}$, density $n^*$=$n\sigma^3$, polarization density $P^*$=$P\sqrt{\epsilon/\sigma^3}$ etc.
). As the temperature is increased, a rapid decrease in $P$ with gradually decreasing $n$ are found, manifesting the Curie (ferroelectric/paraelectric or ferromagnetic/paramagnetic) phase transition. High temperature melt phases for all systems are orientationally disordered with zero $P$, except for the case of $M^*$=2, in which ferroelectric {\it o}melt phase manifests by the sudden increase of $P$ as the $T$ is lowered.  Crystal/melt coexistence properties for the five systems are listed in TabS.\ref{tab1}. %More information about the temperature-dependence of the bulk properties are provided below.

%==============================================================
\vspace{.5cm}
\noindent {\bf III. Supporting Information of the MC Model Prediction of $\mu$}
\vspace{.3cm}
%==============================================================

In the MC (Mikheev and Chernov) model [34] for the fcc/melt CMI, $\mu$ is predicted by the expression as follows,
\begin{equation}
 \tag{S25}
\mu_{\hat{n}}^\mathrm{MC}=\frac{L\xi_{b}}{8k_\mathrm{B} T_{m}^{2}\varsigma_1 A_{\hat{n}}^\mathrm{MC}}
\end{equation}
in which, $\xi_{b}$ is the correlation length of the liquid corresponding to the inverse half-width of the liquid structure factor, $\varsigma_1$ is the first set of dissipative time constant defined in Eq.7 in the Methods. The dimensionless anisotropy factor $A_{\hat{n}}^\mathrm{MC}$ is given by, 
\begin{equation}
 \tag{S26}
A_{\hat{n}}^\mathrm{MC}=\sum_{\hat{K}_{i}} \xi_{b}\int\mathrm{d}z\sum_{\vec{K}_i}(\frac{\mathrm{d}u_i}{\mathrm{d}z})^2
\end{equation}
in which the length scale $\int\mathrm{d}z\sum_{\vec{K}_i}(\frac{\mathrm{d}u_i}{\mathrm{d}z})^2$ is further approximated with the asymptotic exponential decay length from the linear truncated density functional theory, yielding,
\begin{equation}
 \tag{S27}
A_{\hat{n}}^\mathrm{MC}=\frac{1}{8}(\sum_\mathrm{T} (\xi_{b}|\hat{K}_{i}|)^{\frac{1}{2}}+\sum_\mathrm{N T} \frac{1}{\left|\hat{K}_{i} \cdot \hat{n}\right|})
\end{equation}
where ``T'' stands for the transverse density waves (with respect to the CMI orientation $\hat{n}$) with $\hat{K}_{i} \cdot \hat{n}=0$, ``NT'' stands for the density waves with non-zero value of $(\hat{K}_{i} \cdot \hat{n})$. %Based on above expressions, one can find that the fcc/melt CMI kinetic anisotropy predicted by the MC model, e.g., $\mu^\mathrm{MC}_{100}/\mu^\mathrm{MC}_{011}$, is simply governed by the structure property.

$\xi^*_{b}$ for the three fcc/melt CMIs ($M^*$=0, $\frac{1}{2}$, 1) have nearly identical value of 1.79, so the estimated values of $A_{100}^\mathrm{MC}$ and $A_{011}^\mathrm{MC}$ also have constant values 1.7 and 2.4, respectively, over the three $M^*$ systems. Other input parameters in the Eq.S25 are listed in the TABLE.I in the main text. The results of the fcc/melt CMI $\mu$ predicted by the MC model are summarized in TABLE.I together with the NEMD results as a function of dipole strength. It is found that the magnitudes of $\mu$ predicted by the MC model are far from being comparable (five times smaller) with NEMD results, due to its oversimplification of the density waves across the CMI by neglecting nonlinear interactions between density waves\cite{Mikheev91,Wu15}.

%==============================================================
\vspace{.5cm}
\noindent {\bf IV. Supporting Data for the Validation of the TDGL Theory}
%==============================================================

%==============================================================
%\noindent A. RLVs of Crystals and the First-Peak Wavenumber of the Liquid Structure Factors
%==============================================================

As has been noted by other workers\cite{Ramakrishnan79,Wu07,Mikheev91}, the magnitudes of the fcc eight principal RLVs lie very close to $k_1$, $k$=$|\vec{K}_{\langle111\rangle}|$$\approx k_1$, whereas the amplitudes of the second set of RLVs ($k$=$|\vec{G}_{\langle200\rangle}|$) are found to lie away from peaks in the $S(k)$. As seen in Tab.S\ref{tabS5}, this connection between the principal RLV and liquid structure factor holds perfectly well in the cases of $M^*$=0, $\frac{1}{2}$, 1.

\begin{table*}
\renewcommand{\tablename}{Tab.S}
\caption{The magnitudes (in reduced units) of the $|\vec{K}_{\langle111\rangle}|$ and $|\vec{G}_{\langle200\rangle}|$ corresponding to the eight $[111]$ and six $[200]$ RLVs of three fcc crystal system (at the corresponding $T^*_\mathrm{m}$), as well as the magnitudes of $|\vec{K}_{u_a}|$, $|\vec{K}_{u_b}|$, $|\vec{G}_{v_a}|$ and $|\vec{G}_{v_b}|$ corresponding to the principal and the second set RLVs of the {\it o}bco crystal systems (at the corresponding $T^*_\mathrm{m}$). Also listed include the first peak wavenumber values of $k_1$ in the static structure function $S(k)$, and the first peak wavenumber values of $q_1$ in the dipolar structure response function $S_o(q)$. The superscript ``P'' on $k$ and $q$ of the structure factors and the dipolar structure response functions, stands for the component parallel to the polarization direction. The superscripts ``$o_\mathrm{I}$'' and ``$o_\mathrm{m}$'' stand for the polarization density state of the melt phases.}
\begin{ruledtabular}
\begin{tabular}{cccccllll}
$M^*$&\multicolumn{2}{c}{$|\vec{K}^*_{\langle111\rangle}|$}&\multicolumn{2}{c}{$|\vec{G}^*_{\langle200\rangle}|$}&\multicolumn{2}{r}{$k_1^*$}& &\\
\hline
0    &\multicolumn{2}{c}{6.73}&\multicolumn{2}{c}{7.77}& \multicolumn{2}{r}{6.75}& &\\
1/2 &\multicolumn{2}{c}{6.73}&\multicolumn{2}{c}{7.77}& \multicolumn{2}{r}{6.75}& &\\
1    &\multicolumn{2}{c}{6.78}&\multicolumn{2}{c}{7.83}& \multicolumn{2}{r}{6.78}& &\\
\hline
$M^*$&$|\vec{K}^*_{u_a}|$&$|\vec{K}^*_{u_b}|$&$|\vec{G}^*_{v_a}|$&$|\vec{G}^*_{v_b}|$&$k^{\mathrm{P}*}_1[S^{o_{\mathrm{m}}}(k^\mathrm{P})]$&$k^{\mathrm{P}*}_1[S^{o_{\mathrm{I}}}(k^\mathrm{P})]$&$q^{\mathrm{P}*}_1[S_o^{o_{\mathrm{m}}}(q^\mathrm{P})]$&$q^{\mathrm{P}*}_1[S_o^{o_{\mathrm{I}}}(q^\mathrm{P})]$\\
\hline
$\sqrt{2}$&6.79&6.97&7.95&7.92&6.88&7.08&6.88&7.08\\
2            &6.68&7.28&8.16&7.88&7.33&7.33&7.33&7.33\\
\end{tabular}
\end{ruledtabular}
\label{tabS5}
\end{table*} % double column
%\end{sidewaystable}

%==============================================================
%==============================================================
%\vspace{.5cm}
%\noindent B. GL Order Parameter Profiles
%\vspace{.3cm}
%==============================================================

%\begin{widetext}
\begin{figure*}[htbp]
\centering
\includegraphics[width=6.5 in]{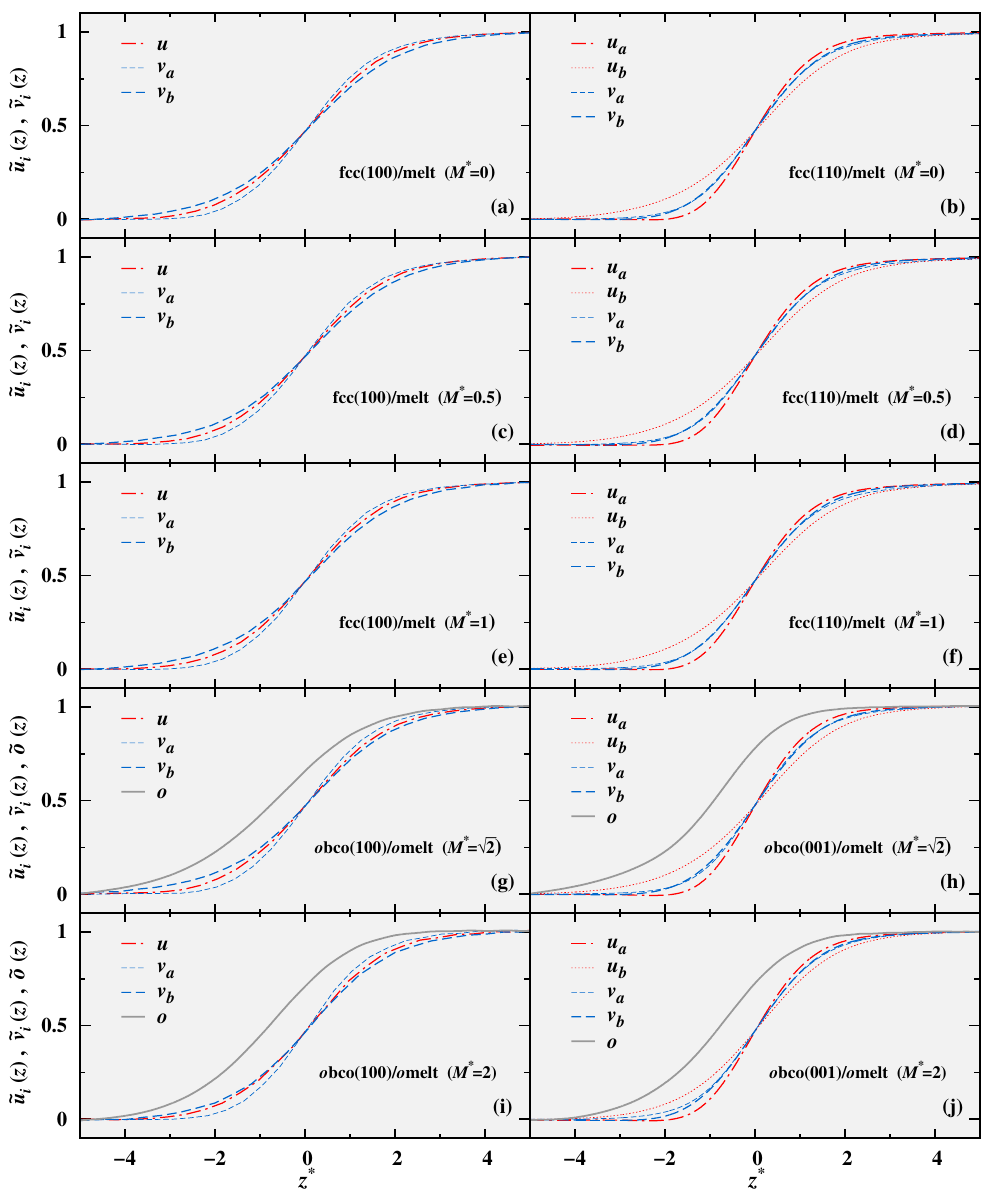}
\renewcommand{\figurename}{Fig. S3}
\renewcommand{\thefigure}{}
\caption{The reduced order parameter profiles for all CMIs computed from the equilibrium MD simulations using Eq.4 and Eq.5 (in the Methods) with the $\vec{K}$ and $\vec{G}$ RLV subsets listed in Tab.S\ref{tabKnGn} and Tab.S\ref{tabKnGn-bco}. The $z$ coordinate is measured relative to $\xi(t)$, $z^*<0$ for melt and $z^*>0$ for crystal.}
\end{figure*}

\begin{widetext}

\begin{table}[]
\renewcommand{\tablename}{Tab.S}
\caption{The bulk values of the GL order parameter extracted from the GL order parameter profiles. $o_\mathrm{I}$ indicates the polarization density for the interface orientationally ordered melt extracted from the $o(z)$ profile.}
\begin{ruledtabular}
\begin{tabular}{cccccccccccc}
\multicolumn{12}{c}{fcc(100)/melt}\\
\hline
$M^*$&$u_{\mathrm{c}}$&$u_{\mathrm{m}}$&$u_{a\mathrm{c}}$&$u_{a\mathrm{m}}$&$v_{b\mathrm{c}}$&$v_{b\mathrm{m}}$&&&&&\\
\hline
0&0.85&0.02&0.80&0.01&0.80&0.02&&&&&\\
$1/2$&0.85&0.02&0.80&0.01&0.80&0.02&&&&&\\
1&0.85&0.02&0.80&0.01&0.80&0.02&&&&&\\
\hline
\multicolumn{12}{c}{fcc(011)/melt}\\
\hline
$M^*$&$u_{a\mathrm{c}}$&$u_{a\mathrm{m}}$&$u_{b\mathrm{c}}$&$u_{b\mathrm{m}}$&$v_{a\mathrm{c}}$&$v_{a\mathrm{m}}$&$v_{b\mathrm{c}}$&$v_{b\mathrm{m}}$&&&\\
\hline
0&0.85&0.03&0.85&0.02&0.80&0.02&0.80&0.01&&&\\
$1/2$&0.85&0.03&0.85&0.02&0.80&0.02&0.80&0.01&&&\\
1&0.85&0.03&0.85&0.02&0.80&0.02&0.80&0.01&&&\\
\hline
\multicolumn{12}{c}{{\it o}bco100/({\it o})melt}\\
\hline
$M^*$&$u_{a\mathrm{c}}$&$u_{a\mathrm{m}}$&$u_{b\mathrm{c}}$&$u_{b\mathrm{m}}$&$v_{a\mathrm{c}}$&$v_{a\mathrm{m}}$&$v_{b\mathrm{c}}$&$v_{b\mathrm{m}}$&$o_{\mathrm{c}}$&$o_{\mathrm{m}}$&$o_{\mathrm{I}}$\\
\hline
$\sqrt{2}$&0.85&0.03&0.85&0.02&0.80&0.01&0.80&0.02&0.98&0.01&0.50\\
2&0.85&0.02&0.82&0.02&0.80&0.01&0.80&0.02&1.79&1.25&1.40\\
\hline
\multicolumn{12}{c}{{\it o}bco001/({\it o})melt}\\
\hline
$M^*$&$u_{a\mathrm{c}}$&$u_{a\mathrm{m}}$&$u_{b\mathrm{c}}$&$u_{b\mathrm{m}}$&$v_{a\mathrm{c}}$&$v_{a\mathrm{m}}$&$v_{b\mathrm{c}}$&$v_{b\mathrm{m}}$&$o_{\mathrm{c}}$&$o_{\mathrm{m}}$&$o_{\mathrm{I}}$\\
\hline
$\sqrt{2}$&0.85&0.03&0.85&0.02&0.80&0.02&0.80&0.01&0.99&0.01&0.50\\
2&0.85&0.03&0.82&0.02&0.80&0.02&0.76&0.01&1.79&1.25&1.40\\
\end{tabular}
\end{ruledtabular}
\end{table} % double column

%==============================================================

%\begin{sidewaystable}[!ht]
\begin{table}[]
\renewcommand{\tablename}{Tab.S}
\caption{Summary of the input parameters for calculating dissipative time constants based on Eq.7, Eq.8 and Eq.12 in the Methods. (i) The density wave relaxation times and the orientational relaxation times, measured from the dynamic structure factors and dipolar structure response functions. (ii) The static structure factors at wavenumbers equal to the magnitudes of RLVs listed in Tab.S\ref{tabS5}, and the static dipolar structure response function at wavenumber of the first peak. It is known that the frequency width at half-max of the dynamic structure factor $S(k,\omega)$ has a minimum at the $k$ value corresponding to the first peak $k_1$ in the static structure function $S(k)$, e.g. shown by the inelastic neutron scattering experiments carried out by Cohen et al.\cite{Cohen87}. For example, $M^*$=0 system, our result is consistent with the findings of inelastic neutron scattering experiments, e.g., $\tau^*(|\vec{K}_{\langle111\rangle}|)$ is 2.7 times greater than $\tau^*(|\vec{G}_{\langle200\rangle}|)$. However, the two DTCs ($\varsigma^*_1=\tau^*(|\vec{K}_{\langle111\rangle}|)/S(|\vec{K}_{\langle111\rangle}|)$ and $\varsigma^*_2=\tau^*(|\vec{G}_{\langle200\rangle}|)/S(|\vec{G}_{\langle200\rangle}|)$) are found to have the identical value, which is a consequence that $S(|\vec{K}_{\langle111\rangle}|)$ is also around 2.7 times greater than $S(|\vec{G}_{\langle200\rangle}|)$.
}
\begin{ruledtabular}
\begin{tabular}{cccccccccccc}
$M^*$&\multicolumn{2}{c}{$\tau^*(|\vec{K}^*_{\langle111\rangle}|)$}&\multicolumn{2}{c}{$S(|\vec{K}^*_{\langle111\rangle}|)$}&\multicolumn{2}{c}{$\tau^*(|\vec{G}^*_{\langle200\rangle}|)$}&\multicolumn{2}{c}{$S(|\vec{G}^*_{\langle200\rangle}|)$} &&&\\
\hline
0&\multicolumn{2}{c}{1.22(3)}&\multicolumn{2}{c}{2.80(5)}&\multicolumn{2}{c}{0.45(3)}&\multicolumn{2}{c}{1.02(4)}&\\
$1/2$&\multicolumn{2}{c}{1.22(4)}&\multicolumn{2}{c}{2.80(5)}&\multicolumn{2}{c}{0.44(2)}&\multicolumn{2}{c}{1.02(3)}&\\
1&\multicolumn{2}{c}{1.25(3)}&\multicolumn{2}{c}{2.80(5)}&\multicolumn{2}{c}{0.50(2)}&\multicolumn{2}{c}{1.00(3)}&&&\\
\hline
$M^*$&$\tau^{o_{\mathrm{m}}*}(|\vec{K}^*_{u_a}|)$&$\tau^{o_{\mathrm{m}}*}(|\vec{K}^*_{u_b}|)$&$\tau^{o_{\mathrm{m}}*}(|\vec{G}^*_{v_a}|)$&$\tau^{o_{\mathrm{m}}*}(|\vec{G}^*_{v_b}|)$&$\tau_o^{o_{\mathrm{m}}*}(q^\mathrm{P}_{1})$&$S^{o_{\mathrm{m}}}(|\vec{K}^*_{u_a}|)$&$S^{o_{\mathrm{m}}}(|\vec{K}^*_{u_b}|)$&$S^{o_{\mathrm{m}}}(|\vec{G}^*_{v_a}|)$&$S^{o_{\mathrm{m}}}(|\vec{G}^*_{v_b}|)$&$S_o^{o_{\mathrm{m}}}(q^\mathrm{P}_{1})$&\\
\hline
$\sqrt{2}$&1.27(2)&1.28(3)&0.53(1)&0.53(2)&0.53(4)&2.57(5)&2.85(7)&1.09(2)&1.09(2)&1.30(7)&\\
$2$&0.66(2)&1.10(4)&0.48(1)&0.56(3)&1.21(5)&1.58(3)&3.18(8)&1.09(2)&1.46(5)&7.1(2)&\\
\hline
$M^*$&$\tau^{o_{\mathrm{I}}*}(|\vec{K}^*_{u_a}|)$&$\tau^{o_{\mathrm{I}}*}(|\vec{K}^*_{u_b}|)$&$\tau^{o_{\mathrm{I}}*}(|\vec{G}^*_{v_a}|)$&$\tau^{o_{\mathrm{I}}*}(|\vec{G}^*_{v_b}|)$&$\tau_o^{o_{\mathrm{I}}*}(q^\mathrm{P}_{1})$&$S^{o_{\mathrm{I}}}(|\vec{K}^*_{u_a}|)$&$S^{o_{\mathrm{I}}}(|\vec{K}^*_{u_b}|)$&$S^{o_{\mathrm{I}}}(|\vec{G}^*_{v_a}|)$&$S^{o_{\mathrm{I}}}(|\vec{G}^*_{v_b}|)$&$S_o^{o_{\mathrm{I}}}(q^\mathrm{P}_{1})$&\\
\hline
$\sqrt{2}$&1.15(3)&1.33(4)&0.59(1)&0.59(1)&1.30(3)&2.41(6)&2.81(5)&1.18(2)&1.18(2)&2.00(8)&\\
$2$&0.58(2)&1.27(5)&0.47(2)&0.63(2)&1.27(6)&1.33(5)&3.60(8)&1.06(2)&1.52(2)&10.1(3)&\\
\end{tabular}
\end{ruledtabular}
\end{table}
%\end{sidewaystable}
% double column

%\subsection{C. Structure factors and dipolar structure response functions, relaxation times}

\begin{figure}[]
\includegraphics[width=5.5 in]{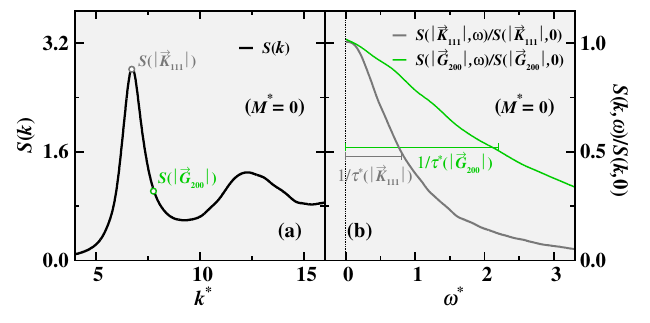}
\renewcommand{\figurename}{Fig. S4}
\renewcommand{\thefigure}{}
\caption{(a) The static structure factor $S(k)$ of the ($M^*=0$) melt phase at $T^*_\mathrm{m}(M^*=0)$. The gray and green points on the $S(k)$ curve correspond to the wavenumber equals $|\vec{K}^*_{\langle111\rangle}|$ and $|\vec{G}^*_{\langle200\rangle}|$, respectively. (b) The density wave relaxation times, $\tau^*(|\vec{K}^*_{\langle111\rangle}|)$ and $\tau^*(|\vec{G}^*_{\langle200\rangle}|)$, are determined from the dynamic structure factors (normalized by their zero frequency values) for the same melt phases in (a), under constant wavenumber $k^*=|\vec{K}^*_{\langle111\rangle}|$ and $k^*=|\vec{G}^*_{\langle200\rangle}|$.}
\end{figure}

\begin{figure}[]
\includegraphics[width=5.5 in]{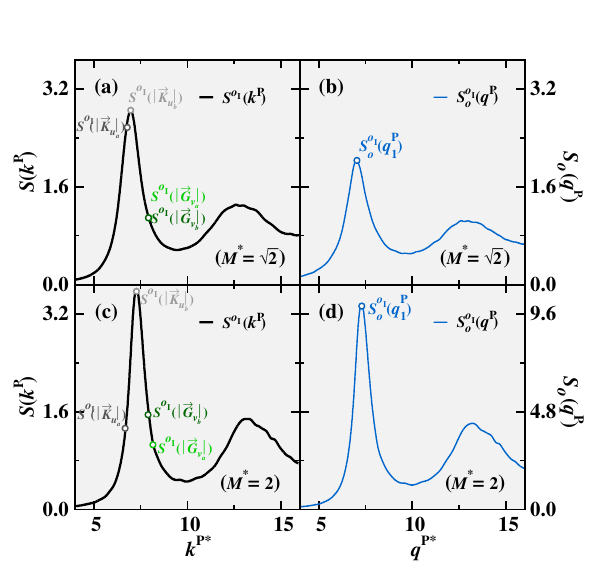}
\renewcommand{\figurename}{Fig. S5}
\renewcommand{\thefigure}{}
\caption{Panel (a) and (c), the static structure factors (component parallel to the polarization direction, ``P'') of the ($M^*$=$\sqrt{2}$, 2) melt phases at polarization density of $o_{\mathrm{I}}$ and $T^*_\mathrm{m}$. The circles on the structure factor curves correspond to the wave numbers with the same magnitudes of RLVs listed in the Table.SII. Panel (b) and (d), the static dipolar structure response functions (component parallel to the polarization direction, ``P'') for the same melt phases in (a) and (c). $S^{o_{\mathrm{I}}*}_{\it o}(q^\mathrm{P}_1)$ is the maximum value at the wavenumber of the first peak $q_1={q^\mathrm{P}_1}$.}
\end{figure}

\begin{figure}[]
\includegraphics[width=5.5 in]{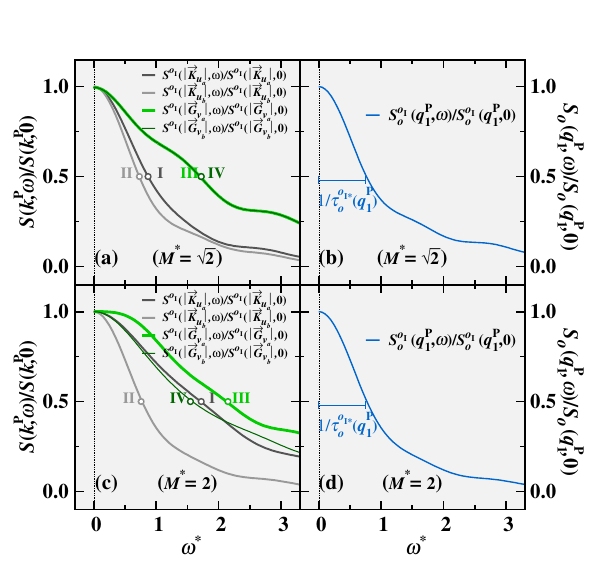}
\renewcommand{\figurename}{Fig. S6}
\renewcommand{\thefigure}{}
\caption{Panel (a) and (c), the normalized dynamic structure factors (component parallel to the polarization direction, ``P'') of the ($M^*$=$\sqrt{2}$, 2) melt phases at polarization density of $o_\mathrm{I}$ and $T^*_\mathrm{m}$. The inverse x-axis values ($1/\omega$) for the circles labeled with ``I'', ``II'', ``III'' and ``IV'' correspond to the density wave relaxation times $\tau^{o_{\mathrm{I}}*}(|\vec{K}_{u_a}|)$, $\tau^{o_{\mathrm{I}}*}(|\vec{K}_{u_b}|)$, $\tau^{o_{\mathrm{I}}*}(|\vec{G}_{v_a}|)$ and $\tau^{o_{\mathrm{I}}*}(|\vec{G}_{v_b}|)$, respectively. Panel (b) and (d), the normalized dynamic dipolar structure response functions (component parallel to the polarization direction, ``P'') for the same melt phases in (a) and (c). The corresponding orientational relaxation time scales, $\tau^{o_{\mathrm{I}}*}_{\it o}$($q^\mathrm{P}_1$) are also labeled out.}
\end{figure}

\end{widetext}

%==============================================================

\newpage

%==============================================================
\vspace{.5cm}
\noindent {\bf V. Supplementary Videos}
%==============================================================

{\bf Movie 1}. Animation of the steady-state growth of fcc crystal of EDM ($M^*=1$) particles along (100) CMI orientation. The undercooling is $\Delta T^*=0.014$. The movie include 500 successive frames (with 1000 MD time steps between neighboring frames) corresponding to a total NEMD simulation length of 500.0 (Lennard-Jones time unit). The color-coding and the viewpoint are set same as Fig.2(a) in the main text. 
%The \href{https://www.icloud.com/iclouddrive/0t8PYKk20n7CggjASF282W-YQ#SM-Movie1}{movie 1} can be downloaded via the hyperlink.

{\bf Movie 2}. Animation of the steady-state growth of orientational ordered bco crystal along {\it o}bco(100) from the orientational disordered melt phase. $M^*=\sqrt{2}$. The undercooling is $\Delta T^*=0.018$. The movie include 500 successive frames (with 1000 MD time steps between neighboring frames) corresponding to a total NEMD simulation length of 500.0 (Lennard-Jones time unit). The color-coding and the viewpoint are set same as Fig.2(b) in the main text. %The \href{https://www.icloud.com/iclouddrive/0lN4_Pij9EEIrpGJ75tK_5NRQ#SM-Movie2}{movie 2} can be downloaded via the hyperlink.

{\bf Movie 3}. Animation of the steady-state growth of orientational ordered bco crystal along {\it o}bco(100) from the orientational ordered melt phase. $M^*=2$. The undercooling is $\Delta T^*=0.023$. The movie include 500 successive frames (with 1000 MD time steps between neighboring frames) corresponding to a total NEMD simulation length of 500.0 (Lennard-Jones time unit). The color-coding and the viewpoint are set same as Fig.2(c) in the main text. %The \href{https://www.icloud.com/iclouddrive/0nEquOHXtV-eSOrjAkJ-E3Ltw#SM-Movie3}{movie 3} can be downloaded via the hyperlink.

%==============================================================
\end{document}